\documentclass[12pt]{article}
\pdfoutput=1

\usepackage{cancel,slashed}
\usepackage{bbm}
\usepackage{amssymb}
\usepackage{amsfonts}
\usepackage{amsmath}
\usepackage{graphicx}
\usepackage{cite}

\usepackage{ dsfont }
\usepackage{latexsym}
\usepackage{color}
\usepackage{xypic}
\usepackage{cancel,slashed}
\usepackage[hyperfootnotes=false,linktocpage]{hyperref}
\usepackage{color}
\usepackage{transparent}
\usepackage[hang,flushmargin]{footmisc}

\usepackage{blkarray}
\usepackage{multirow}

\newcommand{\bmat}{\left(\begin{array}}
\newcommand{\emat}{\end{array}\right)}

\def\Z{\mathbb{Z}}

\def\CK {{\cal K}}
\def\a {\alpha}
\def\b {\beta}

\def\ov{\overline}

\def\IM{\text{Im}\,}
\def\RE{\text{Re}\,}
\def\ov{\overline}
\def\1{{\bf 1}}
\def\2{{\bf 2}}
\def\3{{\bf 3}}
\def\4{{\bf 4}}
\def\6{{\bf 6}}

\def\targ#1#2{\genfrac{[}{]}{0pt}{}{#1}{#2}}
\def\targ2#1#2{\genfrac{}{}{0pt}{}{#1}{#2}}

\definecolor{mygr}{rgb}{0,0.6,0}
\definecolor{mygrey}{rgb}{0,0.1,0.2}
\definecolor{myblue}{rgb}{0,0.5,0.9}
\definecolor{myblue2}{rgb}{0,0.5,0.5}
\definecolor{myblue3}{rgb}{0,0.7,0.9}
\definecolor{myblue4}{rgb}{0,0.6,0.6}
\definecolor{myorange}{rgb}{1,0.5,0}
\definecolor{mypurple}{rgb}{0.6,0,1}
\definecolor{mygolden}{rgb}{1,0.8,0.2}
\definecolor{mycyan}{rgb}{0,1,1}
\definecolor{mymagenta}{rgb}{1,0,1}
\definecolor{mykiwi}{rgb}{0.8,1,0.5}
\definecolor{mybrown}{cmyk}{0.14, 0.42, 0.56, 0.2}
\definecolor{myturq}{cmyk}{0.99, 0, 0.2, 0.4}
\definecolor{myaubergine2}{cmyk}{0.4, 0.5, 0, 0.1}
\definecolor{myaubergine}{cmyk}{0.6,0.85,0,0}
\definecolor{CycleGreen}{cmyk}{0.52,0,1,0}
\definecolor{CycleBrown}{cmyk}{0, 0.4, 0.9, 0.2}

\DeclareFontFamily{U}{rcjhbltx}{}
\DeclareFontShape{U}{rcjhbltx}{m}{n}{<->rcjhbltx}{}
\DeclareSymbolFont{hebrewletters}{U}{rcjhbltx}{m}{n}

\DeclareMathSymbol{\lamed}{\mathord}{hebrewletters}{108}
\DeclareMathSymbol{\mem}{\mathord}{hebrewletters}{109}
\DeclareMathSymbol{\ayin}{\mathord}{hebrewletters}{96}
\DeclareMathSymbol{\tsadi}{\mathord}{hebrewletters}{118}
\DeclareMathSymbol{\qof}{\mathord}{hebrewletters}{113}
\DeclareMathSymbol{\resh}{\mathord}{hebrewletters}{114}
\DeclareMathSymbol{\pe}{\mathord}{hebrewletters}{112}
\DeclareMathSymbol{\pesofit}{\mathord}{hebrewletters}{80}
\DeclareMathSymbol{\samekh}{\mathord}{hebrewletters}{115}
\DeclareMathSymbol{\tav}{\mathord}{hebrewletters}{116}
\DeclareMathSymbol{\vav}{\mathord}{hebrewletters}{119}
\DeclareMathSymbol{\het}{\mathord}{hebrewletters}{120}
\DeclareMathSymbol{\yod}{\mathord}{hebrewletters}{121}
\DeclareMathSymbol{\zayin}{\mathord}{hebrewletters}{122}
\DeclareMathSymbol{\alephdot}{\mathord}{hebrewletters}{128}
\DeclareMathSymbol{\tsadisofit}{\mathord}{hebrewletters}{90}
\DeclareMathSymbol{\shin}{\mathord}{hebrewletters}{152}

\def\CN {{\cal N}}
\def\trh {{\tilde{\rho}}}
\def\sig{{\sigma}}
\def\d{{\delta}}
\def\be{\begin{equation}}
\def\ee{\end{equation}}
\def\bea{\begin{eqnarray}}
\def\eea{\end{eqnarray}}
\def\bes{\begin{subequations}}
\def\ees{\end{subequations}}
\def\raw{\rightarrow}

\def\eps{{\epsilon}}
\def\oh{\frac{1}{2}}
\def\CG {{\cal G}}

\usepackage[normalem]{ulem}

\usepackage{multicol}
\usepackage{float}
\usepackage{caption}
\def\mk {{\mathcal K}}

\def\r {{\rho}}
\def\hr {{\hat{\rho}}}
\def\tr {{\tilde{\rho}}}
\def\p {{\partial}}
\def\e {{\epsilon}}
\def\g {{\gamma}}
\def\hg {{\hat{\gamma}}}
\def\te {{\tilde{\epsilon}}}
\def\he {{\hat{\epsilon}}}
\def\tg {{\tilde{\gamma}}}

\def\s {{\sigma}}

\def\th {{\theta}}
\def\hth {{\hat{\theta}}}
\def\ph   {{\phi}}

\newcommand{\cK}{\mathcal{K}}
\newcommand{\cM}{\mathcal{M}}
\newcommand{\cN}{\mathcal{N}}

\makeatletter
\newsavebox\myboxA
\newsavebox\myboxB
\newlength\mylenA

\newcommand*\xoverline[2][0.75]{%
\sbox{\myboxA}{$\m@th#2$}%
\setbox\myboxB\null
\ht\myboxB=\ht\myboxA%
\dp\myboxB=\dp\myboxA%
\wd\myboxB=#1\wd\myboxA
\sbox\myboxB{$\m@th\overline{\copy\myboxB}$}
\setlength\mylenA{\the\wd\myboxA}
\addtolength\mylenA{-\the\wd\myboxB}%
\ifdim\wd\myboxB<\wd\myboxA%
   \rlap{\hskip 0.5\mylenA\usebox\myboxB}{\usebox\myboxA}%
\else
    \hskip -0.5\mylenA\rlap{\usebox\myboxA}{\hskip 0.5\mylenA\usebox\myboxB}%
\fi}
\makeatother

\begin{document}
\pagestyle{plain}

\makeatletter
\@addtoreset{equation}{section}
\makeatother
\renewcommand{\theequation}{\thesection.\arabic{equation}}

\pagestyle{empty}
\rightline{IFT-UAM/CSIC-19-113}
\vspace{0.5cm}
\begin{center}
\Huge{{A Landscape of AdS Flux Vacua}
\\[10mm]}
\large{Fernando Marchesano and Joan Quirant \\[10mm]}
\small{
Instituto de F\'{\i}sica Te\'orica UAM-CSIC, Cantoblanco, 28049 Madrid, Spain
\\[8mm]} 
\small{\bf Abstract} \\[5mm]
\end{center}
\begin{center}
\begin{minipage}[h]{15.0cm} 

We analyse type IIA Calabi-Yau orientifolds with background fluxes and D6-branes. Rewriting the F-term scalar potential as a bilinear in flux-axion polynomials yields a more efficient description of the Landscape of flux vacua, as they are invariant under the discrete shift symmetries of the 4d effective theory. In particular, expressing the extremisation conditions of the scalar potential in terms of such polynomials allows for a systematic search of vacua. We classify families of $\cN=0$ Minkowski, $\cN=1$ AdS and $\cN=0$ AdS flux vacua, extending previous findings in the literature to the Calabi-Yau context. We compute the spectrum of flux-induced masses for some of them and show that they are perturbatively stable, and in particular find a branch of $\cN=0$ AdS vacua where tachyons are absent. Finally, we extend this Landscape to the open string sector by including mobile D6-branes and their fluxes. 

\end{minipage}
\end{center}
\newpage
\setcounter{page}{1}
\pagestyle{plain}
\renewcommand{\thefootnote}{\arabic{footnote}}
\setcounter{footnote}{0}


\tableofcontents

\section{Introduction}
\label{s:intro}

A fundamental question in the context of string compactifications is the characterisation of the string Landscape, that is the collection of isolated, metastable 4d vacua that are obtained from string theory. In this regard compactifications with background fluxes \cite{Grana:2005jc,Douglas:2006es,Becker:2007zj,Denef:2007pq,Ibanez:2012zz} have proven to be a remarkably fruitful framework. To great extent, this is because they provide a simple mechanism for moduli stabilisation that at the same time generates a discretum of vacua, which allows developing our intuition on how the full string Landscape may look like.

Within the flux landscape, a very interesting corner is given by (massive) type IIA flux compactifications, in the sense that one may achieve full moduli stabilisation using only classical ingredients. Early results on this subject display a non-trivial set of classical IIA flux vacua to AdS$_4$ \cite{Behrndt:2004km,Behrndt:2004mj,Derendinger:2004jn,Lust:2004ig,Villadoro:2005cu,DeWolfe:2005uu,House:2005yc,Camara:2005dc,Koerber:2007jb,Aldazabal:2007sn,Tomasiello:2007eq,Koerber:2008rx,Caviezel:2008ik,Lust:2008zd,Cassani:2009ck,Lust:2009zb,Lust:2009mb,Koerber:2010rn,Narayan:2010em}. Some of these solutions are based on the results of \cite{Grimm:2004ua}, which combines the classical K\"ahler potential of Calabi-Yau (CY) orientifolds and the superpotential induced by RR and NS background $p$-form fluxes to obtain an effective F-term potential. In particular, ref.\cite{DeWolfe:2005uu} obtains a discretum of $\CN=1$ AdS$_4$ vacua from such an effective 4d approach. The same strategy was implemented in \cite{Camara:2005dc} for the specific case in which the Calabi-Yau is a six-torus, finding different branches of supersymmetric and non-supersymmetric AdS$_4$ vacua. 

In this paper we extend the general analysis of \cite{DeWolfe:2005uu} to find further vacua of the classical 4d potential of \cite{Grimm:2004ua}, which are not necessarily supersymmetric. The motivation to analyse this particular setup is two-fold: on the one hand, it has been recently shown in \cite{Herraez:2018vae} that the type IIA CY flux potential can be expressed as a bilinear on the flux quanta, in which the dependence of axions and saxions factorises. As such, the extremisation conditions take a particularly simple form, already exploited in \cite{Escobar:2018tiu,Escobar:2018rna} in the search for new vacua. On the other hand, Calabi-Yau orientifolds with fluxes constitute an interesting arena to test the recent Swampland conjectures involving string compactifications to AdS \cite{Ooguri:2016pdq,Lust:2019zwm}, and in principle they could provide counterexamples to them. In order to properly address whether or not this is the case, it is important to determine the full set of vacua that corresponds to this construction. 

Needless to say, solving for general vacua of a potential is more involved than restricting the search to supersymmetric ones. In the last case, even when the K\"ahler metrics for moduli fields are not fully specified, the vanishing conditions for the F-terms allow rewriting the vacua conditions algebraically, significantly simplifying the analysis. Interestingly, the factorised form of the potential found in \cite{Herraez:2018vae}, which features a number of flux-axion polynomials invariant under discrete shift symmetries, allows implementing  a similar strategy in the search of more general vacua. Indeed, we find that by imposing a simple off-shell Ansatz for the derivatives of the potential in terms of the flux-axion polynomials, the extrema conditions can also be expressed algebraically. By solving them we find several branches of extrema, one of which corresponds to supersymmetric AdS vacua, other to the non-supersymmetric Minkowski vacua discussed in \cite{Palti:2008mg}, and the rest are different branches of non-supersymmetric AdS solutions. Compared to previous results in the literature, on the one hand we find a one-to-one correspondence between our branches of solutions and those found in \cite{Camara:2005dc} for isotropic toroidal compactifications. On the other hand, we find that some of the extrema found in \cite{Narayan:2010em} are incompatible with our results. Our approach also permits to analyse the perturbative stability of these new AdS solutions, solving for the spectrum of flux-induced masses for the simplest branches of extrema. In those cases we find some branches where tachyons are absent, and some others where they are present but satisfy the Breitenlohner-Freedman bound. Finally, our strategy can be easily generalised to include moduli and fluxes in the open string sector, providing an even richer landscape of AdS flux vacua. 

The rest of the paper is organised as follows. In section \ref{s:IIAorientifold} we review the setup of type IIA Calabi-Yau orientifolds with fluxes, the classical F-term potential associated to them and its bilinear formulation. In section \ref{s:IIAvacua} we implement our Ansatz to solve for the extrema conditions, finding several branches of solutions which are summarised in table \ref{vacuresul}. In section \ref{s:stability} we analyse the perturbative stability of some of these branches and find that they can be considered perturbatively stable, see table \ref{tablee}. Section \ref{s:validity} discusses the validity of these solutions from both a 4d and a 10d viewpoint. Section \ref{s:D6branes} generalises the setup to include D6-brane with moduli and the corresponding worldvolume fluxes. We draw our conclusions in section \ref{s:conclu}, and relegate some technical details to the appendices. Appendix \ref{ap:relations} contains some K\"ahler metric relations used in the main text, while appendix \ref{ap:Hessian} performs a detailed analysis of the Hessian for several branches of solutions.


\section{Type IIA orientifolds with fluxes}
\label{s:IIAorientifold}

Type IIA flux compactifications constitute a very interesting sector of the string landscape, in the sense that from the classical flux potential one obtains both 4d Minkowski and AdS vacua, some with all moduli stabilised  \cite{Grana:2005jc,Douglas:2006es,Ibanez:2012zz}. In the following we will focus on (massive) type IIA flux vacua whose internal geometry can be approximated by a Calabi-Yau orientifold, as assumed in \cite{Grimm:2004ua} to derive the F-term potential used in \cite{DeWolfe:2005uu}.\footnote{Using such potential to search for vacua is justified a posteriori, by arguing that the flux-induced scale can be made parametrically smaller than the Kaluza-Klein scale, in the same region where corrections to the potential can be neglected, see \cite{DeWolfe:2005uu} and section \ref{s:validity4d}. Therefore, even if in the presence of fluxes the compactification metric is not Calabi-Yau, it is expected that the fluxless K\"ahler potential is a good approximation to capture the 4d dynamics. See also \cite{McOrist:2012yc,Gautason:2015tig} for some objections to this approach.} We then express the scalar potential in the factorised bilinear form of \cite{Herraez:2018vae}. As pointed out in there, the bilinear form of the potential is independent on whether the background geometry is Calabi-Yau or not and, as it will be clear from the computations in the next section, so will be the strategy to extract the vacua from it.

\subsection{Type IIA on Calabi-Yau orientifolds}

Let us consider type IIA string theory compactified on an orientifold of $\mathbb{R}^{1,3} \times \cM_6$ with $\cM_6$ a compact Calabi-Yau three-fold. More precisely, we take the standard orientifold quotient by $\Omega_p (-)^{F_L} {\cal R}$ \cite{Blumenhagen:2005mu,Blumenhagen:2006ci,Marchesano:2007de,Ibanez:2012zz},\footnote{Here $\Omega_p$ stands for worldsheet parity and  $(-)^{F_L}$ for a projection operator counting the number of spacetime fermions in the left-moving sector.}  with ${\cal R}$ an anti-holomorphic Calabi-Yau involution acting on the K\"ahler 2-form $J$ and the holomorphic 3-form $\Omega$ as ${\cal R}(J) = - J$ and ${\cal R} (\Omega) = \ov \Omega$, respectively.

In the absence of background fluxes, and neglecting worldsheet and D-brane instanton effects, dimensional reduction to 4d will yield several massless chiral fields, whose scalar components can be described as follows \cite{Grimm:2004ua}. On the one hand, we have the complexified K\"ahler moduli $T^a = b^a + it^a$ defined through 
\begin{equation}
J_c \equiv B + i\, e^{\frac{\phi}{2}} J = \left( b^a + i t^a\right) \omega_a \, , \qquad \quad a \in \{1, \ldots, h^{1,1}_- \},
\end{equation} 
where $J$ is expressed in the Einstein frame and $\phi$ represents the ten-dimensional dilaton. The 2-form basis $\ell_s^{-2}\omega_a$ correspond to harmonic representatives of the classes in $H^{2}_-({\cal M}_6, \Z)$ and are dimensionless due to the insertion of the string length $\ell_s = 2 \pi \sqrt{\alpha'}$. The kinetic terms for these moduli is encoded in their K\"ahler potential
\be
K_K \,  = \, -{\rm log} \left(\frac{i}{6} \CK_{abc} (T^a - \bar{T}^a)(T^b - \bar{T}^b)(T^c - \bar{T}^c) \right) \, = \,  -{\rm log} \left(\frac{4}{3} \cK\right) \, ,
\label{KK}
\ee
where ${\cal K}_{abc} =  \ell_s^{-6} \int_{{\cal M}_6} \omega_a \wedge \omega_b \wedge \omega_c$ are the Calabi-Yau triple intersection numbers and $\cK = \cK_{abc} t^at^bt^c = 6 {\rm Vol}_{\cM_6} = \frac{3}{4} {\cal G}_T$ is a homogeneous function of degree three on the $t^a$.

On the other hand, the complex structure moduli of the compactification pair up with the axions arising from RR three-form potential $C_3$ as follows. One first defines the complexified 3-form $\Omega_c$ as
\begin{equation}
\Omega_c \equiv C_3 + i \, \RE ({\cal C} \Omega)\, ,
\end{equation}
where ${\cal C} \equiv  e^{- \phi} e^{\frac{1}{2}(K_{cs} - K_T)}$ is a compensator, with $K_{cs} = - \log \left(- i\ell_s^{-6} \int_{{\cal M}_6} \Omega \wedge  \ov \Omega \right)$. Then one takes a symplectic basis $(\alpha_\kappa, \beta^\lambda) \in H_3({\cal M}_6, \Z)$ such that the holomorphic three-form can be written as $\Omega = {\cal Z}^\kappa \alpha_\kappa - {\cal F}_\lambda \beta^\lambda$. The orientifold projection  decomposes this basis into ${\cal R}$-even $(\alpha_K, \beta^\Lambda) \in H_+^3({\cal M}_6,\Z)$ and ${\cal R}$-odd 3-forms $(\beta^K, \alpha_\Lambda) \in H_-^3({\cal M}_6,\Z)$, and eliminates half of the degrees of freedom of the original complex periods of $\Omega$. Finally, the complex structure moduli are defined in terms of the ${\cal R}$-odd 3-form basis:
\begin{equation}\label{cpxmoduli}
N^K = \xi^K + i n^K = \ell_s^{-3} \int_{{\cal M}_6} \Omega_c \wedge \beta^K, \qquad  U_{\Lambda} = \xi_\Lambda + i u_\Lambda =  \ell_s^{-3} \int_{{\cal M}_6} \Omega_c \wedge \alpha_\Lambda.
\end{equation}
Their kinetic terms are given in terms of the following piece of the K\"ahler potential:
\begin{equation}\label{KQ}
 K_Q = -2 \log \left( \frac{1}{4} \RE({\cal C}{\cal Z}^K) \IM({\cal C} {\cal F}_K) - \frac{1}{4} \IM({\cal C} {\cal Z}^\Lambda) \RE({\cal C} {\cal F}_\Lambda) \right) = -  \log(e^{-4 D}),
\end{equation}
where $D$ is the four-dimensional dilaton defined through $e^{D} \equiv \frac{e^{\phi}}{\sqrt{{\rm Vol}_{\cM_6}}}$. The periods ${\cal F}_K$ and ${\cal F}_\Lambda$ ought to be considered as homogeneous functions of degree one in the periods ${\cal Z}^K$ and ${\cal Z}^\Lambda$, implying that the function ${\cal G}_Q= e^{-K_Q/2}$ is a homogeneous function of degree two in $n^K$ and $u_{\Lambda}$. The complex structure moduli \eqref{cpxmoduli} are redefined in the presence of D6-brane moduli, and so is the K\"ahler potential \eqref{KQ}. For simplicity, we will not consider this case for now, leaving its discussion to section \ref{s:D6branes}.

\subsection{The type IIA flux potential}

On top of the above orientifold background one may add RR and NS background fluxes. One may describe them in terms of the democratic formulation of type IIA supergravity \cite{Bergshoeff:2001pv}, in which all RR potentials are grouped in a polyform ${\bf C} = C_1 + C_3 + C_5 + C_7 + C_9$, and so are their field strengths ${\bf G} = G_0 + G_2 + G_4 + G_6 + G_8 + G_{10}$. The Bianchi identities for such field strengths read
\begin{equation}\label{IIABI}
\ell_s^2\, d (e^{B} \wedge {\bf G} ) = - \sum_\a \delta (\Pi_\alpha) \wedge e^{-F_\alpha}, \qquad d H = 0 \, ,
\end{equation} 
where we have also included the BI for NS flux $H$. Here $\Pi_\alpha$ hosts a localised source with a worldvolume flux $F_\alpha$, and $\delta(\Pi_\alpha)$ is the bump $\delta$-function form with support on $\Pi_\alpha$ and indices transverse to it.  The solution to \eqref{IIABI} can then be decomposed as
\begin{equation}\label{IIABIsol}
{\bf G} = e^{-B}\wedge (d {\bf A} + \ov{\bf G})\,, \qquad H = dB + \ov H\, ,
\end{equation}
where ${\bf A} = {\bf C}\wedge e^{B}$ and $\ov{\bf G}$ is a sum of closed $p$-forms to be thought as the background values for the internal RR fluxes. One may now impose Page charge quantisation \cite{Marolf:2000cb},
\begin{equation}
\frac{1}{\ell_s^{2p-1}} \int_{\pi_{2p}} d A_{2p-1} + \ov G_{2p} \in \Z, \qquad  \frac{1}{\ell_s^{2}}\int_{\pi_3} dB + \ov H \in \Z,
\end{equation} 
where $\pi_{2p}$ with $p=1,2,3$ and $\pi_3$ are internal cycles of $\cM_6$. In the absence of localised sources such as D-branes, the gauge potentials~${\bf A}$ are well-defined everywhere and the cohomology class of $\ov{G}_{2p}$, $\ov H$ along $\cM_6$ capture  the internal flux quanta. For orientifold compactifications the internal $p$-cycles have to comply with the orientifold projection, such that    
the background flux can be characterised by virtue of flux quanta $(m,m^a,e_a,e_0)$. These are defined as
\begin{equation} 
\ell_s {\ov G}_0 = -m, \qquad \frac{1}{\ell_s} \int_{\tilde \pi^a}{\ov G}_2 = m^a, \qquad \frac{1}{\ell_s^3}\int_{\pi_a} {\ov G}_4 = -e_a, \qquad  \frac{1}{\ell_s^5}\int_{{\cal M}_6} {\ov G}_6 = e_0,
\end{equation}   
with $[\pi_a] \in H_4^{+}({\cal M}_6, \Z)$ Poincar\'e dual to $[\ell_s^{-2} \omega_a]$, and  $[\tilde \pi^a] \in H_2^{-}({\cal M}_6, \Z)$ Poincar\'e dual to $[\ell_s^{-4} \tilde{\omega}^a]$, where $\ell_s^{-6}\int_{X_6} \omega_a \wedge \tilde{\omega}^b = \delta^b_a$. The internal RR-fluxes ${\ov{\bf G}}$ are known to generate a perturbative superpotential for the K\"ahler moduli \cite{Gukov:1999ya,Taylor:1999ii}:
\begin{equation} \label{WT}
\ell_s W_T = \frac{1}{\ell_s^5} \int_{{\cal M}_6} {\ov{\bf G}} \wedge e^{ -J_c} = e_0 + e_a T^a + \frac{1}{2} {\cal K}_{abc} m^a T^b T^c + \frac{m}{6} {\cal K}_{abc} T^a T^b T^c\, .
\end{equation}
The NS 3-form flux ${\ov H}_3$ on the other hand threads the ${\cal R}$-odd three-cycles $(B^K,A_\Lambda) \in H^-_3({\cal M}_6, \Z)$, which are the de Rham duals to the ${\cal R}$-odd three-forms $(\beta^K, \alpha_\Lambda)$ introduced earlier. Similar as for the RR-fluxes, the quantised Page charge for the NS-flux background can be expressed in terms of the integer flux quanta $(h_K, h^\Lambda)$:
\begin{equation}\label{Hflux}
 \frac{1}{\ell_s^2} \int_{B^K} {\ov H} = h_K\, ,\qquad   \frac{1}{\ell_s^2} \int_{A_\Lambda} {\ov H} = - h^\Lambda \,,
\end{equation}
and generate a linear superpotential for the complex structure moduli
\begin{equation} \label{WQ}
\ell_s W_{Q} = \frac{1}{\ell_s^5} \int_{{\cal M}_6} \Omega_c \wedge {\ov H}_3 = h_K N^K  + h^\Lambda U_{\Lambda}\, . 
\end{equation}
The combination of RR and NS-fluxes suffices to generate a four-dimensional F-term scalar potential for the geometric moduli $(t^a, n^K, u_{\Lambda})$ and closed string axions $(b^a,\xi^K, \xi_{\Lambda})$, whose precise shape exhibits a remarkable factorisation into a geometric moduli piece, an axion piece and a flux piece~\cite{Bielleman:2015ina,Herraez:2018vae}. Namely, we have a bilinear structure of the form
\begin{equation}\label{VF}
V  = \frac{1}{\kappa_4^2} \vec{\rho}^{\ t}\, {\bf Z} \, \vec{\rho}\, , 
\end{equation}  
where the matrix ${\bf Z}$ only depends on the saxions $\{t,n,u\}$, while the vector $\vec{\rho}$ only on  the flux quanta and the axions $\{b, \xi\}$. More precisely, the dependence of the flux quanta is linear, and so one may write $\ell_s\vec{\rho}  = R' \cdot \vec{q}$, with $R'$ an axion-dependent rotation matrix and $\vec{q} = (e_0, e_a, m^a, m, h_K,  h^\Lambda)^t$ the vector of flux quanta. In general the entries of $\vec{\rho}$ are axion polynomials with flux-quanta coefficients that are invariant under the discrete shift symmetries of the combined superpotential $W = W_T + W_Q$. In the case at hand they read
 \begin{equation}\label{rhos}
\begin{array}{lcl}
\ell_s \rho_0 &=& e_0 + e_a b^a + \frac{1}{2} {\cal K}_{abc} m^a b^b b^c + \frac{m}{6} {\cal K}_{abc} b^a b^b b^c + h_\mu \xi^\mu, \\
\ell_s \rho_a &=& e_a + {\cal K}_{abc}  m^b b^c + \frac{m}{2} {\cal K}_{abc} b^b b^c, \\
\ell_s \tilde \rho^a &=& m^a + m b^a , \\
\ell_s \tilde \rho &=& m\,, \\
\ell_s \hat \rho_\mu & = & h_\mu\,,
\end{array}
\end{equation}
where for simplicity we have gathered the NS fluxes as $h_\mu = (h_K, h^\Lambda)$, and similarly for the complex structure fields $\xi^\mu = (\xi^K, \xi_\Lambda)$, $u^\mu = (n^K, u_\Lambda)$. In this basis, the saxion-dependent matrix ${\bf Z}$ reads
\begin{equation}\label{ZAB}
{\bf Z} = e^{K} \left(\begin{array}{ccccc} 
4 & \\
& K^{a b} \\
&&\frac{4}{9} {\cal K}^2 K_{a b} \\
&&& \frac{1}{9} {\cal K}^2 & \frac{2}{3} {\cal K} u^\mu \\ 
&&&  \frac{2}{3} {\cal K} u^\nu & K^{\mu\nu}  
 \end{array} \right)\, ,
\end{equation} 
where $K = K_K + K_Q$, $K_{ab} = \frac{1}{4} \partial_{t^a} \partial_{t^b} K_K$, and $K_{\mu\nu} = \frac{1}{4} \partial_{u^\mu}\partial_{u^\nu} K_Q$, and with upper indices denote their inverses. As shown in \cite{Escobar:2018rna}, this structure is maintained when including the effect of curvature $\alpha'$-corrections. The same is true in the presence of D6-brane moduli, as discussed in \cite{Escobar:2018tiu}  and reviewed in section \ref{s:D6branes}.


\section{Type IIA flux vacua}
\label{s:IIAvacua}

As already exploited in \cite{Escobar:2018tiu,Escobar:2018rna}, the bilinear structure of  F-term potential \eqref{VF} can be used to look for vacua in type IIA flux compactifications. In this section we will generalise this approach and implement a quite general strategy for the search of extrema of $V$, that will lead to different branches of solutions for the case of CY orientifold flux backgrounds. These branches will mostly describe new non-supersymmetric AdS solutions, but they will also contain the supersymmetric AdS solutions of \cite{DeWolfe:2005uu} and the non-supersymmetric Minkowski solutions of \cite{Palti:2008mg}. As we will see, these vacua correspond to the branches of the toroidal type IIA flux vacua found in \cite{Camara:2005dc}, but now generalised to the much broader context of Calabi-Yau geometries. In the next section we will analyse the  spectrum of some of these extrema and see that they are, in fact, classically stable AdS vacua.

\subsection{Extrema conditions}

Let us start by writing explicitly the different extrema conditions, grouped into the first order derivatives of the F-term potential \eqref{VF} with respect to the axions $\{\xi^\mu, b^a\}$ and saxions $\{u^\mu, t^a\}$ of the compactification. Using the explicit expressions for {\bf Z} and $\vec{\rho}$ we find:

\vspace*{.5cm}

\textbf{Axionic directions}

\bes
\label{paxions}
\begin{equation}
\label{paxioncpx}
\left.\frac{\partial V}{\partial \xi^\mu}\right|_{\rm vac}  = \left. 8 e^K\rho_0 \hat{\rho}_\mu\right|_{\rm vac} =0
\end{equation}
\begin{equation}
\label{paxionk}
\left.\frac{\partial V}{\partial b^a}\right|_{\rm vac}  = e^K\left[ 8 \rho_0\rho_a+\frac{8}{9}\mathcal{K}^2\tilde{\rho}^cK_{ca}\tilde{\rho}+ 2\rho_c K^{cd}\mathcal{K}_{dla}\tilde{\rho}^l\right]_{\rm vac}=0 
\end{equation}
\ees

\vspace*{.5cm}

\textbf{Saxionic directions}

\bes
\label{psaxions}
\begin{equation}
\label{psaxioncpx}
\left.\frac{\partial V}{\partial u^\mu}\right|_{\rm vac} = e^K\left[e^{-K} V \partial_\mu K+\frac{4}{3}\mk\tr\hr_\mu +\partial_\mu K^{\kappa\sigma}\hr_\kappa\hr_\sigma\right]_{\rm vac}=0
\end{equation}
\begin{equation}
\label{psaxionk}
\left.\frac{\partial V}{\partial t^a}\right|_{\rm vac} = e^K\left[e^{-K} V\partial_{a}K+\partial_{a}\left(\frac{4}{9}\mathcal{K}^2\tilde{\rho}^b\tilde{\rho}^c
 K_{bc}\right)+\partial_{a}K^{cd}\rho_c\rho_d+\mathcal{K}_a\tilde{\rho}\left(\frac{2}{3}\mathcal{K}\tilde{\rho}+4u^\mu\hat{\rho}_\mu \right)\right]_{\rm vac}=0
\end{equation}
\ees

Interestingly, manipulating these condition one may rederive the inequality found in \cite{Hertzberg:2007wc} that in turn prevents the existence of de Sitter vacua. Indeed, using the properties listed in appendix \ref{ap:relations} it is straightforward to see that, off-shell: 
\begin{align}
\label{dss}
 u^\mu\partial_{u^\mu} V+\frac{1}{3}t^a\partial_{t^a} V =-3V-\frac{8e^K}{27}\mathcal{K}^2\tilde{\rho}^a\tilde{\rho}^bK_{ab}-8e^K\rho_0^2-\frac{4e^K}{3}K^{ab}\rho_a\rho_b\, .
\end{align}
At each extremum, where $\p V=0$, this equation shows that $V|_{\text{extremum}}$ must be negative or vanishing. In particular at a vacuum $V|_{\text{vac}}\leq 0$, forbidding any dS vacuum at the classical level. It would be interesting to see if the above kind of relation is preserved or violated by the different corrections to the classical approximation, along the lines of \cite{Banlaki:2018ayh}.

\subsection{The Ansatz}
\label{ss:ansatz}

Rather than solving the extrema conditions \eqref{paxions} and \eqref{psaxions} by brute force, in the following we will use the algebraic properties of the axion polynomials $\rho_A$ to set up an Ansatz to look for vacua. To describe such Ansatz, we will first convert the vector $\vec{\rho}$ into a different vector $\vec{\gamma}$, of the form
\be
\vec{\rho} \ \raw \ \vec{\g} \, =\, 
\left(
\begin{array}{c}
\g_0  \\ \g_a \\ \tilde{\g}^a \\ \hat\g_\mu  \\ \tilde{\rho} 
\end{array}
\right)\, =\, 
\left(
\begin{array}{c}
\rho_0 -  \tilde{\rho} \epsilon_0  \\ \rho_a - \tilde{\rho} \eps_a \\ \tilde{\rho}^a - \tilde{\rho} \tilde{\eps}^a  \\ \hat\rho_\mu - \tilde{\rho} \hat\eps_\mu \\ \tilde{\rho}
\end{array}
\right) \, ,
\label{Hbasis}
\ee
which can be seen as a (field-dependent) change of basis. The moduli-dependent functions $\eps$ are such that $\vec{\gamma}$ has only one non-vanishing component at the vacuum. Namely we define them such that $\vec{\gamma}|_{\rm vac} = ( 0 \quad 0 \quad 0 \quad 0 \quad  \tilde{\rho})^t$. Of course, this does not really constrain what the $\epsilon$'s may be, because there is an infinite number of functions with the same value at a single point. However, we will impose an Ansatz that will significantly constrain this freedom. Indeed, in the following we will look at vacua such that, off-shell,
\begin{equation}
\boxed{\partial_\alpha V\, =\, \chi_\alpha^A \gamma_A}\, 
\label{Ansatz}
\end{equation}
where $\gamma_A = \{\gamma_0, \gamma_a, \tilde{\gamma}^a, \hat{\gamma}_\mu \}$ runs over all the components of $\vec{\gamma}$ except $\tilde{\rho}$, and $\chi_\alpha^A$ are some regular functions of the moduli, with the latter indexed by $\alpha$. Notice that this essentially implies that the $\epsilon_A$ are also regular functions of the moduli.

In order to implement this Ansatz, it proves useful to rewrite the potential and the extrema conditions in terms of the $\vec{\gamma}$ basis. We have that
\begin{equation}
    V  = \frac{1}{\kappa_4^2} \vec{\gamma}^{\ t}\, \hat{\bf Z} \, \vec{\gamma},
\end{equation}
where, unlike the $\rho_A$, the elements of $\vec{\g}$ are regular functions that depend on both the axions and the saxions. The bilinear product now reads
\be
\hat{\bf Z} = 
e^{K} \left(
\begin{array}{c  c  c  c c}
4 & & & & \eps_0  \\
 & K^{ab} & & & K^{ab}\eps_a \\
& & \frac{4}{9}\CK^2K_{ab} & & \frac{4}{9}\CK^2K_{ab}\tilde{\eps}^a \\
& & & K^{\mu\nu} &  \frac{2}{3} \CK u^\nu +  K^{\mu\nu} \hat\eps_\mu \\
\eps_0 & K^{ab}\eps_b &  \frac{4}{9}\CK^2K_{ab}\tilde{\eps}^b  &  \frac{2}{3} \CK u^\mu +  K^{\mu\nu} \hat\eps_\nu  & \frac{1}{9}\CK^2 + \a
\end{array}
\right)\, ,
\label{Zg}
\ee
where
\be
\a = \eps_0^2 + K^{ab}\eps_a\eps_b + \frac{4}{9}\CK^2K_{ab}\tilde{\eps}^a\tilde{\eps}^b + K^{\mu\nu}\hat\eps_\mu \hat\eps_\nu+ \frac{4}{3} \CK u^\mu \hat\eps_\mu\, .
\ee
The strategy will now be to extremise $V$ in this basis, in order to obtain the different expressions for the $\eps$'s or, in other words, the functional dependence of $\g_A$. To each class of solutions will correspond a different class of vacua. 

Notice that we can split the scalar potential as
\be\label{Vsplit}
V = V_1 + V_2 =  \vec{\g}^{\, t}\,  \hat{\bf Z}_{1} \, \vec{\g} +  \vec{\g}^{\, t}\,  \hat{\bf Z}_{2} \, \vec{\g}\, ,
\ee
where
\be
\hat{\bf Z}_{1} = 
e^{K} \left(
\begin{array}{c  c  c  c c}
4 & & &  \\
 & K^{ab} & & &  \\
& & \frac{4}{9}\CK^2K_{ab} & &  \\
& & & K^{\mu\nu} &  \\
& & & & 0
\end{array}
\right)\, ,
\label{Zg1}
\ee
and
\be
\hat{\bf Z}_{2} = 
e^{K} \left(
\begin{array}{c  c  c  c c}
0 & & & & \eps_0 \\
 & 0 & & & K^{ab}\eps_a \\
& & 0 & & \frac{4}{9}\CK^2K_{ab}\tilde{\eps}^a \\
& & & 0 &  \frac{2}{3} \CK u^\nu +  K^{\mu\nu} \hat\eps_\mu \\
\eps_0 & K^{ab}\eps_b &  \frac{4}{9}\CK^2K_{ab}\tilde{\eps}^b  &  \frac{2}{3} \CK u^\mu +  K^{\mu\nu} \hat\eps_\nu  & \frac{1}{9}\CK^2 + \a
\end{array}
\right)\, .
\label{Zg2}
\ee
Note also that $V_1$ is positive semidefinite, while $V_2$ is not. Because $V_1$ is quadratic on quantities that vanish at the vacuum, the extremisation conditions are equivalent to taking derivatives with respect to $V_2$ only
\be
\p V|_{\rm vac} = 0 \quad \iff \quad \p V_2|_{\rm vac} = 0\, .
\label{extrema}
\ee
In this sense, our Ansatz \eqref{Ansatz} requires something stronger than \eqref{extrema}. Namely that, off-shell, $\p V_2$ is a function which is at least linear in the $\g_A$. In the following we will classify the different classes of solutions that arise from this requirement. 

\subsection{Branches of vacua} 
\label{ss:branches}

Let us now turn to solve for the extrema conditions \eqref{paxions} and \eqref{psaxions}. As we will see, rewriting them in the form \eqref{extrema} makes it easier to classify the different branches of solutions. Later on we will discuss how such branches reproduce and generalise previous vacua found in the literature. 


\subsubsection*{Axionic derivatives}

Already from the initial expression \eqref{VF}, \eqref{rhos}, \eqref{ZAB}, one can see that $V$ depends quadratically on $\rho_0$, which is the only quantity that depends on the complex structure axions $\xi^\mu$. Moreover, as it depends linearly we have that 
\begin{equation}
\label{rho0}
\partial_{\xi_\mu} V=  8 e^K\rho_0 \frac{\partial \rho_0}{\partial \xi_\mu}= 8e^K\rho_0 \hat{\rho}_\mu\, , \qquad \partial_{\xi_\mu} V|_{\rm vac} =0  \rightarrow \boxed{\rho_0|_{\rm vac}=0}\, .
\end{equation}
Therefore, in our Ansatz \eqref{Hbasis} one may take $\eps_0 \equiv 0$, as we will do in the following. 

Let us now look at the derivative with respect to the B-field axions:
\be
\p_{b^a} V_2 =  \tilde{\rho}^2 e^K \left[ \frac{8}{9}\CK^2K_{ab} \tilde{\eps}^b+2 K^{bd}\eps_d \CK_{abc}\tilde{\epsilon}^c \right]+ \dots
\label{bax}
\ee
where we have used that $\p_{b^a} \rho_b = \CK_{abc}\tilde{\rho}^c$ and $\p_{b^a} \tilde\rho^b = \tilde{\rho} \d_a^b$, and the dots stand for terms linear in the $\gamma_A$. The Ansatz \eqref{Ansatz} has then two possible solutions:
\begin{itemize}
\item \textbf{Branch A1:} 
\begin{align}
&\boxed{\tilde\epsilon^b=0}	&	\rightarrow&	& \tilde{\rho}^b\rvert_{\rm vac}&=0\, .
\end{align}

\item \textbf{Branch A2:} 

Let us assume that $\tilde\epsilon^b\neq 0$ and multiply \eqref{bax} by $t^a$. Using the relations in appendix \ref{ap:relations}, one sees that a necessary condition for the bracket in the rhs of \eqref{bax} to vanish off-shell is
\begin{align}
&\boxed{\epsilon_d=-\frac{1}{4}\mathcal{K}_d	}	&	&\rightarrow 	&	\rho_a\rvert_{\rm vac}=&-\frac{1}{4}\tilde{\rho}\mathcal{K}_a\, .
\end{align}
Replacing this result in \eqref{bax} one obtains a 2nd condition:
\begin{align}
&\boxed{\tilde \epsilon^a=Bt^a}	&	&\rightarrow	&	\tilde{\rho}^a\rvert_{\rm vac}&=Bt^a\, ,
\end{align}
with $B\neq0$ some regular function of the moduli.

\end{itemize}



\subsubsection*{Saxionic derivatives}

The saxionic derivatives conditions are, for the complex structure moduli:
\begin{align}
&\p_{u^\sig} V_2 =\trh^2e^{K}\left[\p_{u^\sig} K \left( \frac{\CK^2}{9} + \a\right) + \left( \p_{u^\sig} K^{\mu\nu}\right) \hat\eps_\mu \hat\eps_\nu + \frac{4}{3} \CK  \hat\eps_\sig \right]+\nonumber\\&+\p_{u^\sig}\left(\frac{4e^K}{3}\mathcal{K}\tilde{\rho}u^\mu+2e^K\tilde{\rho} K^{\mu \nu } \hat{\epsilon }_{\nu }\right)\hat{\gamma}_\mu+\p_{u^\sig}\left(\frac{8e^K}{9}\mathcal{K}^2\tilde{\rho}K_{ba}\tilde{\epsilon}^b\right)\tilde{\gamma}^a+\p_{u^\sig}\left(2e^K\tilde{\rho }K^{ab} \epsilon _a\right) \gamma_b\, .
\label{cpxsax}
\end{align}
Notice that if one contracts \eqref{cpxsax} with $u^\sig$ and uses that $u^\sig \p_{u^\sig} K^{\mu\nu} = 2 K^{\mu\nu}$ one obtains:
\be
-\frac{ e^{-K}}{4\trh^2}u^\sig\p_{u^\sig} V_2= \oh K^{\mu\nu} \hat\eps_\mu \hat\eps_\nu + \CK u^\mu  \hat\eps_\mu+ \left(\frac{1}{9}\CK^2 + K^{ab}\eps_a\eps_b + \frac{4}{9}\CK^2K_{ab}\tilde{\eps}^a\tilde{\eps}^b\right)+\dots \, 
\label{trcpxsax}
\ee
where the dots stand for terms linear in $\g_A$.

Finally, the K\"ahler saxionic derivative reads:
\begin{align}
&\p_{t^a} V_2 =e^K\trh^2 \left[\p_{t^a} K \left( \frac{\CK^2}{9} + \a\right) +\frac{1}{9} \p_{t^a} \CK^2 + \left( \p_{t^a} K^{bc}\right) \eps_b\eps_c + \frac{4}{9}\p_{t^a} \left( \CK^2K_{ab}\right) \tilde{\eps}^a\tilde{\eps}^b + {4} \CK_a  u^\mu \hat\eps_\mu\right] +\nonumber\\ &\p_{t^a}\left(\frac{4e^K}{3}\mathcal{K}\tilde{\rho}u^\mu+2e^K\tilde{\rho} K^{\mu \nu } \hat{\epsilon }_{\nu }\right)\hat{\gamma}_\mu+\p_{t^a}\left(\frac{8e^K}{9}\mathcal{K}^2\tilde{\rho}K_{bc}\tilde{\epsilon}^b\right)\tilde{\gamma}^c+\p_{t^a}\left(2e^K\tilde{\rho }K^{cb} \epsilon _c\right)\gamma_b\, .
 \label{kahsax}
\end{align}
Proceeding as before, one can contract  \eqref{kahsax} with $t^a$ to obtain:
\be
\frac{e^{-K}}{\trh^2} t^a\p_{t^a} V_2=\frac{1}{3} \CK^2  - K^{bc} \eps_b\eps_c + \frac{4}{9} \CK^2K_{ab} \tilde{\eps}^a\tilde{\eps}^b  - 3 K^{\mu\nu}\hat\eps_\mu \hat\eps_\nu+\dots\, 
\label{trkahsax}
\ee
where again the dots stand for terms linear in the $\g_A$ and we have used that $t^a  \p_{t^a} K^{bc} = 2  K^{bc}$ and  $t^a \p_{t^a} \left( \CK^2K_{ab}\right) = 4 \CK^2K_{ab}$. 
Notice that both the first line of \eqref{cpxsax} and of \eqref{kahsax} depend on $\tilde{\rho}$ but not on any other component of $\vec{\rho}$. As such, they cannot depend on the $\g_A$. Following our strategy, we will then demand them to vanish off-shell, ensuring our Ansatz \eqref{Ansatz} and therefore that $\p_{u^\sig} V|_{\rm vac} = \p_{t^a} V|_{\rm vac} =0$.

To proceed, let us consider the general Ansatz for $\hat{\eps}_\mu$:
\be\label{ansatz}
\hat{\eps}_\mu = A \CK \p_{u^\mu} K +\mk \hat{\eps}_\mu^{\rm p} \quad {\rm with}\quad    u^\mu \hat{\eps}_\mu^{\rm p} = 0, 
\ee
where $A$ is some function of the moduli, and the factor of $\CK$ has been introduced for later convenience. The term $\hat{\eps}_\mu^{\rm p}$ is a `primitive' component of $\hat{\eps}_\mu$. We will first consider the case where $\hat{\eps}_\mu^{\rm p} = 0$, which we dub:
\begin{itemize}
\item \textbf{Branch S1:} \boxed{\hat{\eps}_\mu^{\rm p} = 0}

On the one hand the vanishing of \eqref{trcpxsax} becomes 
\be
4A -8 A^2 = \frac{1}{9} + \CK^{-2} K^{ab}\eps_a\eps_b + \frac{4}{9} K_{ab}\tilde{\eps}^a\tilde{\eps}^b ,
\label{ptrcpxsax}
\ee
which we impose off-shell. On the other hand the vanishing of \eqref{trkahsax} reads
\be
48 A^2  = \frac{1}{3}   - \CK^{-2} K^{bc} \eps_b\eps_c + \frac{4}{9} K_{ab} \tilde{\eps}^a\tilde{\eps}^b ,
\label{ptrkahsax}
\ee
to be understood also off-shell. Combining these two equations we find
\bea
K_{ab}\tilde{\eps}^a\tilde{\eps}^b& =&  - \frac{1}{2} + 9A  + 45 A^2  ,\\
 \frac{K^{ab}\eps_a\eps_b}{\CK^2} & = & \frac{1}{9}  +2A - 28A^2  .
 \label{quickcheck}
\eea

For the {\bf Branch A1} one finds the following solutions:
\bea\label{A1S1}
A = \frac{1}{15} & \raw & \boxed{\eps_a = \pm \frac{3}{10}\CK_a}\,  \boxed{\hat{\eps}_\mu = \frac{\CK}{15} \p_\mu K} \,\\
A = - \frac{1}{6} & \raw & \eps_a = \pm i \sqrt{\frac{3}{4}}\CK_a\, ,
\eea
the second one being unphysical. For the {\bf Branch A2} one finds
\bea\label{A2S1}
A = \frac{1}{12} & \raw & B^2 = \frac{1}{4} \raw \boxed{\tilde \epsilon^a=\pm \oh t^a} \,  \boxed{\hat{\eps}_\mu = \frac{\CK}{12} \p_\mu K}\, \\
A = - \frac{1}{84} & \raw & B^2 < 0 \, ,
\eea
again the second solution being unphysical. 

\item \textbf{Branch S2:} \boxed{\hat{\eps}_\mu^{\rm p} \neq 0}

Finding solutions in this branch is in general more involved, as one needs some more specific information on the K\"ahler potential for the dilaton and complex structure moduli. Things however simplify if one considers a K\"ahler potential of the form
\be
K_Q = - {\rm log} (2s) - 2 {\rm log}\left( \tilde{\CG} (u^i) \right)\, ,
\label{KS2}
\ee
where $\tilde{\CG}$ is a homogeneous function of degree $3/2$ on the geometric complex structure moduli. This kind of K\"ahler potential was used in \cite{Palti:2008mg,Escobar:2018tiu,Escobar:2018rna} to construct $\CN=0$ Minkowski flux vacua. Since in this case the metric for the dilaton and other complex structure moduli decouple, it is natural to make the following Ansatz
\be
\hat{\eps}_0 = E_0 \CK \p_s K = - E_0\frac{\CK}{s}\, , \quad \quad \hat{\eps}_i = E \CK \p_{u^i} K = - 2E \CK \frac{\p_i \tilde{\CG}}{\tilde{\CG}}\, ,
\ee
with $E$, $E_0$ functions of the moduli. Then we may easily derive two equations from \eqref{cpxsax}, namely
\bea
\p_s V_2 = 0 & \raw & 8  E_0^2\CK^2 - \frac{4}{3} E_0\CK^2 =  \left( \frac{1}{9}\CK^2 + \a\right) \, , \\
u^i \p_{u^i} V_2 = 0 & \raw & 8 E^2 \CK^2 - \frac{4}{3} E\CK^2 = \left( \frac{1}{9}\CK^2 + \a\right)\, .
\label{nptrcpxsax}
\eea
Notice that $E$, $E_0$ are solutions to the same quadratic equation, so if $E \neq E_0$ then necessarily
\be
E + E_0 = \frac{1}{6}\, .
\label{sumofAs}
\ee
Using this we can rewrite \eqref{nptrcpxsax} as
\be
- 8 E^2  + \frac{8}{3} E = \CK^{-2} K^{ab}\eps_a\eps_b + \frac{4}{9} K_{ab}\tilde{\eps}^a\tilde{\eps}^b .
\ee
Moreover, from  \eqref{trkahsax} and using \eqref{sumofAs} one obtains
\be
48 E^2 - 4 E  = - \CK^{-2} K^{bc} \eps_b\eps_c + \frac{4}{9} K_{ab} \tilde{\eps}^a\tilde{\eps}^b  .
\ee
To sum up, one finds the equations
\bea
K_{ab}\tilde{\eps}^a\tilde{\eps}^b& =& 9 \left(5 E^2 - \frac{1}{6} E\right)  \, ,\\
 \frac{K^{ab}\eps_a\eps_b}{\CK^2} & = & - 28E^2 + \frac{10}{3} E \, .
\eea
In the following we will analyse the possible solutions for the two axionic branches.

For the {\bf Branch A1} one finds the following solutions:
\bea
\label{solMink}
E = 0 & \raw & \boxed{\eps_a = 0}\, , \ \boxed{\hat{\eps}_0 = -\frac{\CK}{6s}}\, , \ \boxed{\hat{\eps}_i = 0}  \, \\
E = \frac{1}{30} & \raw &  \boxed{\eps_a = \pm \frac{ \sqrt{6}}{10}\CK_a} \,  \ \boxed{\hat{\eps}_0 = -\frac{2\CK}{15s}}\, \ \boxed{\hat{\eps}_i = \frac{\CK}{30} \p_{u^i}K} \, .
\label{A1S2}
\eea
One can check that \eqref{solMink} corresponds to the Minkowski vacua analysed in \cite{Palti:2008mg,Escobar:2018tiu}.

For the {\bf Branch A2} one finds
\bea
\label{solS2S1}
E = \frac{1}{12} & \raw & {\tilde\eps^a = \pm \oh t^a},\quad {\hat{\eps}_0 = \frac{\CK}{12}\p_{u^0}K},\quad {\hat{\eps}_i = \frac{\CK}{12} \p_{u^i}K}  \, \\
E = \frac{1}{28} & \raw &  \boxed{\tilde\eps^a = \pm \frac{ 1}{14} t^a} \,  \ \boxed{\hat{\eps}_0 = -\frac{11\CK}{84s}}\,  \ \boxed{\hat{\eps}_i = \frac{\CK}{28} \p_{u^i}K} \, .
\label{A2S2}
\eea
Note that \eqref{solS2S1} is in fact a special case of the Branch S1. For all the other solutions one can express things in terms of the Ansatz \eqref{ansatz} as
\be
\hat{\eps}_\mu = \left(\frac{E}{2} + \frac{1}{24}\right) \CK \p_{u^\mu} K + \mk\hat{\eps}_\mu^{\rm p} \, ,
\ee
with
\be
\hat{\eps}_0^{\rm p} = \left( \frac{1}{8} - \frac{3E}{2}\right)  \p_{s} K\, , \quad \quad \hat{\eps}_i^{\rm p} = \left(\frac{E}{2} - \frac{1}{24}  \right) \p_{u^i} K\, .
\ee

So in total we find two (double) classes of AdS solutions in the Branch {\bf S1} and two (double) classes of AdS solutions in the Branch {\bf S2}, where in the latter we have assumed the factorised metric Ansatz \eqref{KS2}. 

\end{itemize}

\textbf{Uniqueness of the solutions}

Some comments are in order regarding the uniqueness of these solutions. An implicit assumption of the above discussion is that the K\"ahler metric $K_K^{ab}$ is irreducible. If the metric display a block-diagonal structure, as for instance in toroidal orientifolds, then more solutions are recovered. Indeed, one can check that in that case the choice of sign for the $\epsilon_a$'s in \eqref{A1S1} and \eqref{A1S2} can be made independently on each block. Each choice corresponds in principle to a different solution, as it is related to different signs of the flux quanta. The election of the signs will not be reflected in the value of the $V_{\text{vacuum}}$ - which is invariant - but it will affect the F-terms and the spectrum of light modes. Unless stated differently, in the following we will consider a generic irreducible K\"ahler metric, for which the choice of sign must be equal for all $\eps_a$'s.

\subsection{Summary of the vacua and physical properties}

\label{ss:summary}

Let us recap the previous results and compute some of the properties of these extrema:

\textbf{General structure}

All the solutions found for the vacuum equations satisfy:
\begin{align}
\rho_0&=0\, ,		&	\hat{\rho}_\mu &= \trh \CK \left(A \p_{u^\mu} K + \hat{\eps}_\mu^{\rm p}\right) \, ,	& 	\tilde{\rho}^a&=B\tilde\rho t^a\, ,	&	\rho_a&=C\tilde\rho\CK_a\, ,
\label{solutions}
\end{align}
with $A,B,C \in \mathds{R}$. The \textbf{Branch A1} has $B=0$, whereas the \textbf{Branch A2} has $B\neq 0$, $C=-1/4$. The \textbf{Branch S1} has $ \hat{\eps}_\mu^{\rm p}=0$ whereas the \textbf{Branch S2} has $ \hat{\eps}_\mu^{\rm p}\neq0$. It is convenient to point out that as long as $A\neq0$, $C\neq 0$ - ignoring the complex structure axions for the moment - there are as many equations as moduli so in principle all the moduli can be fixed. Regarding the complex structure axions, only the linear combination that appears in the superpotential \eqref{WQ} is fixed. As pointed out in \cite{Camara:2005dc}, this allows the remaining axions to participate in the St\"uckelberg mechanism present in the presence of space-time-filling D6-branes, while guaranteeing the gauge invariance of the flux superpotential.

\bigskip

\textbf{K\"ahler moduli stabilisation}

The structure \eqref{solutions} provides several relations between the K\"ahler moduli and the axion polynomials of the compactification. In particular, the last two equations involving $\rho_a$ and $\tilde{\rho}^a$ provide $2 h^{1,1}_-$ relations between the quantised zero-, two- and four-form fluxes and the complexified K\"ahler moduli. By using \eqref{rhos} one may derive an explicit relation between the geometric K\"ahler moduli and the quantised fluxes. Namely we have that
\begin{equation}
 \hat{e}_a \equiv e_a - \frac{1}{2} \frac{\CK_{abc}m^am^b}{m} = \ell_s \tr \CK_a \left(C - \frac{1}{2} B^2 \right)\, ,
 \label{Kstab}
\end{equation}
where we have defined a shifted four-form flux $\hat{e}_a$ analogous to the one in \cite{DeWolfe:2005uu}, invariant under discrete shifts involving K\"ahler axions and fluxes. It follows from this relation that whenever $B^2 = 2C$ one needs to impose $\hat{e}_a = 0$ in order to have a sensible solution for the extrema conditions, and that then the individual K\"ahler moduli are not stabilised. One can check that this is the case for the branch  \eqref{solMink},  corresponding to the non-supersymmetric Minkowski solutions analysed in \cite{Palti:2008mg}, see also \cite{Escobar:2018tiu,Escobar:2018rna}. As pointed out in there, for Minkowski vacua the constraint on the fluxes $\hat{e}_a = 0$ is lifted once that $\alpha'$ corrections for the K\"ahler sector are taken into account.

\bigskip

\textbf{Vacuum energy}

Using the expressions \eqref{dss} and \eqref{solutions} it is straightforward to see that the vacuum energy has the following general expression:
\begin{equation}
\Lambda= V\rvert_{\rm vac}=-\left(\frac{2}{27}B^2+\frac{16}{27}C^2\right)\frac{e^K}{\kappa_4^{2}}\CK^2\trh^2\, .
\end{equation}

\bigskip

\textbf{F-terms}

Using \eqref{solutions} and the expression for the F-terms derived in \cite{Escobar:2018tiu} one can directly compute them for each of the above extrema
\begin{equation}
F_{T^a}=\trh\CK_a\left(-\frac{C}{2}-\frac{1}{4}+6A\right)+i\CK_a\trh\frac{B}{4}\, ,
\end{equation}
\begin{equation}
F_{U^\mu}=\trh\CK\p_{u^\mu}K\left(\frac{C}{2}-\frac{1}{12}-A+i\frac{B}{4}\right)+\mk\trh\hat\epsilon^p_\mu\, .
\end{equation}

\bigskip
\textbf{Summary}

Finally, we gather all the above results in table \ref{vacuresul}:

\begin{table}[H]
\begin{center}
\scalebox{1}{%
    \begin{tabular}{| c || 	c | c | c | c |c |c |}
    \hline
  Branch & $A$  & $B$  & $C$  & $\kappa_4^{2}\Lambda$  &$F_{T^a}$ & $F_{U^\mu}$\\
  \hline \hline
  \textbf{A1-S1}  &$\frac{1}{15}$ &  $0$  & $\frac{3}{10}$  & $-\frac{4e^K}{75}\CK^2\trh^2$ &0	&0\\ \hline
    \textbf{A1-S1}  &$\frac{1}{15}$  & $0$  & $-\frac{3}{10}$  & $-\frac{4e^K}{75}\CK^2\trh^2$ &$\frac{3\trh}{10}\CK_a$	&$-\frac{3\CK\trh}{10}\p_{u^\mu}K$\\ 
\hline
  \textbf{A1-S2}   & $\frac{7}{120}$  & $0$  & $\pm\frac{\sqrt{6}}{10}$  & $-\frac{8e^K}{225}\CK^2\trh^2$ 	&	$\left(-\frac{C}{2}+\frac{1}{10}\right)\trh\CK_a$&
   {$\begin{aligned}
         F_S&=\left(-\tfrac{1}{15}+\tfrac{C}{2}\right)\CK\trh\p_{s}K\\
          F_{U^i}&=\left(-\tfrac{1}{6}+\tfrac{C}{2}\right)\CK\trh\p_{u^i}K
      \end{aligned}$}  \\ \hline
     \textbf{A1-S2}    & $\frac{1}{24}$ & $0$  & $0$  & $0$ 	&0	& $F_S=0$,\quad  $F_{U^i}=-\frac{\mathcal{K}\tilde{\rho}}{6}\p_{u^i}K$\\ 
\hline
     \textbf{A2-S1}    & $\frac{1}{12}$  & $\pm\frac{1}{2}$  & $-\frac{1}{4}$  & $-\frac{e^K}{18}\CK^2\trh^2$ 	&$\left(\frac{3}{8}+\frac{iB}{4}\right)\tilde{\rho}\mathcal{K}_a$	& $\left(-\frac{7}{24}+\frac{i}{4}B\right)\CK \trh\p_{u^\mu}K$ \\ 
 \hline
      \textbf{A2-S2}    & $\frac{5}{84}$ & $\pm\frac{1}{14}$  & $-\frac{1}{4}$  & $-\frac{11e^K}{294}\CK^2\trh^2$	&$\frac{13}{56}\trh\CK_a$	& 
      {$\begin{aligned}
         F_S&=\left(-\tfrac{11}{56}+\tfrac{iB}{4}\right)\CK\trh\p_{s}K\\
          F_{U^i}&=\left(-\tfrac{7}{24}+\tfrac{iB}{4}\right)\CK\trh\p_{u^i}K
      \end{aligned}$}  \\
      \hline
    \end{tabular}}      
\end{center}
\caption{Different branches of solutions with the corresponding vacuum energy and F-terms. The solutions in the branch {\bf S2} assume the K\"ahler potential \eqref{KS2}. \label{vacuresul}}
\end{table}
As already mentioned, when the structure of the metric in the K\"ahler sector is block diagonal, this allows to choose the sign of $C$ independently in each block and therefore the corresponding value of the F-term. In particular, in the \textbf{Branch A1-S1} one can then break SUSY independently in each of the block-diagonal sectors.

\subsection{Relation to previous results}

As a cross-check of formalism and the solutions discussed so far, let us compare them with some of the existing results in the literature. We will analyse three different papers, presenting their main results schematically. We refer the reader to the original papers for further details. 

\begin{enumerate}
\item \textbf{Comparison with DGKT} \cite{DeWolfe:2005uu}

This paper analyses the general conditions for $\cN=1$ Calabi-Yau orientifold vacua, which are then applied to the particular orbifold background  $\otimes^3_{j=1}T_j^2/\mathds{Z}_3^2$. At the general level, one can easily map our conditions for the {\bf A1-S1} SUSY branch with the equations of section 4 of \cite{DeWolfe:2005uu}. For instance, the condition
\begin{align}\label{DGKT1}
\tr^a=0 \rightarrow b^a =-\frac{m^a}{m}\, ,
\end{align}
is equivalent\footnote{There are some signs differences which arise form the different conventions in the flux quanta definitions.} to (4.33) in  \cite{DeWolfe:2005uu}. This implies that
\begin{align}\label{DGKT2}
\rho_a&=\frac{3}{10}\tilde\rho\CK_a\longrightarrow \hat{e}_a=\frac{3}{10}m\mk_{abc}t^bt^c ,
\end{align}
which is equivalent to (4.36) in \cite{DeWolfe:2005uu}. Regarding the dilaton/complex structure sector, on the one hand one can see that the equations (4.24) and (4.25) in \cite{DeWolfe:2005uu} are equivalent to (3.37) of \cite{Escobar:2018tiu} and to the second condition in \eqref{A1S1}. On the other hand, one can check that the eq.(4.26) of \cite{DeWolfe:2005uu} that fixes one linear combination of axions $\xi$ is equivalent to $\rho_0=0$. 

The same statements hold when applying the above to the specific background $\otimes^3_{j=1}T_j^2/\mathds{Z}_3^2$. Before the inclusion of fluxes, the moduli space of this compactification consists of the axio-dilaton and 12 complexified K\"ahler moduli: 3 of them inherited form the toroidal geometry, and 9 associated with the blow-ups of the orbifold singular points. Since there are no complex structure moduli, the only necessary inputs to solve our equations are the intersection numbers, given by:
\begin{align}
\mk_{ijk}&=\kappa \iff i\neq j\neq k\, ,	&	\mk_{AAA}&=\beta\, ,
\end{align}
where $i,j...$ label the toroidal K\"ahler moduli and $A,B...$ the blow-up modes. Applying \eqref{DGKT1} and \eqref{DGKT2} to this model one finds 
\begin{align}
\tr_i=&\frac{3}{10}\tr\mk_i\rightarrow t_i=\sqrt{\frac{5\hat{e}_j\hat{e}_k}{3m\kappa\hat{e}_i}}\, , 	&	\tr_A&=\frac{3}{10}\tr\mk_A\rightarrow t_A=\sqrt{\frac{10\hat{e}_A}{3\beta m}} \, ,
\end{align}
with $\hat{e}_i=e_i-\kappa\frac{m_jm_k}{m}$, $\hat{e}_A=e_A-\beta\frac{e_A^2}{2m}$, which is equivalent to (5.5) and (5.8) in  \cite{DeWolfe:2005uu}. One can equally recover eqs.(5.10) and (5.12) from applying the conditions of the {\bf A1-S1} SUSY branch. Therefore our results reproduce the analysis in \cite{DeWolfe:2005uu}, as expected.

\item \textbf{Comparison with NT}\cite{Narayan:2010em}

This paper considers the same orbifold background as \cite{DeWolfe:2005uu}, but searches for non-supersymmetric vacua as well. By approximating the potential to its leading terms in certain flux quotients, more solutions to the extremisation equations are found, which are labelled as $\{\text{Case 1), \dots , Case 8)} \}$. Case 1) stands for the supersymmetric solutions already found in \cite{DeWolfe:2005uu}. Case 2) is related to Case 1) by an overall sign flip in all the RR four-form fluxes, that is by an overall sign flip in the $\rho_a$ or equivalently in the $\eps_a$. Therefore, Case 1) and 2) correspond to the two components of the branch \textbf{A1-S1} in table \ref{vacuresul}. Finally, Cases 3), \dots , 8) are obtained by partial sign flips in the four-form fluxes corresponding to the toroidal and blow-up two-cycles, and some of these cases are identified as classically stable vacua while others are not. 

However, one can check that once that the blow-up moduli are introduced the metric in the K\"ahler sector is irreducible. Therefore, from the viewpoint of our analysis, none of the cases 3), \dots , 8) would be actual extrema of the scalar potential. This can be seen for instance by means of the equation \eqref{quickcheck}: performing partial sign flips in the $\epsilon_a$'s will change the LHS for an irreducible K\"ahler metric, while the RHS remains invariant. The fact that the analysis in \cite{Narayan:2010em} identifies these cases as extrema is presumably due to the approximations made in the potential, which effectively removes the kinetic mixing between the different K\"ahler modes.

\item \textbf{Comparison with CFI}\cite{Camara:2005dc}

In this case the CY orientifold is  given by $\otimes^3_{j=1}T_j^2/\Omega_p\left(-1\right)^{F_L}\sigma$, so there are three complexified K\"ahler moduli,  three complex structure moduli and the axio-dilaton. To find different branches of vacua the simplification $T_1=T_2=T_3=T$ is imposed in the K\"ahler sector. The relevant data to apply our results are: 
\begin{align}
\mk_{ijk}&=1 \iff i\neq j\neq k\, ,		&			K_Q&\sim -\log\left(u_0u_1u_2u_3\right)\, ,
\end{align}
where we are using $i,j...$ to label the K\"ahler moduli and $\mu,\nu...$ to label the complex structure moduli ($U^i$) and the axio-dilaton ($U^0)$.
The two branches \textbf{A1} and \textbf{A2} become:
\begin{align}
\tilde{\rho}^b\rvert_{\rm vac}&=0\rightarrow b=-\frac{c_2}{\trh}\, ,	&	\rho_a\rvert_{\rm vac}=&-\frac{1}{4}\tilde{\rho}\mathcal{K}_a\rightarrow b=\frac{-c_2\pm\sqrt{\Gamma-\trh^2t^2/2}}{\trh}\, ,
\end{align}
respectively. Here, as in \cite{Camara:2005dc}, we have dropped the indices in the K\"ahler sector, renamed $m^a=c_2$, $e_a=c_1$ and defined $\Gamma= c_2^2-mc_1$. Notice that these are precisely the two branches found in eq.(4.23) of \cite{Camara:2005dc}, up to some sign due to different conventions in defining flux quanta. Inside each branch, we have distinguished between the subbranches  \textbf{S1} and \textbf{S2} that read:
\begin{align}
\hat{\eps}_\mu^{\rm p}& = 0\rightarrow \hat\rho_k u_k=\hat\rho_0 u_0\, ,	&	E+ E_0 &= \frac{1}{6}\rightarrow \hat\rho_k u_k=\trh t^3-\hat\rho_0u_0 \, ,
\end{align}
which are precisely  the two sub-branches in eq.(4.24) of \cite{Camara:2005dc}. Once that we have matched the branches, is direct to see that, in the vacuum:
\begin{itemize}
\item Branch \textbf{A1}-\textbf{S1}
\begin{align}
\hat\rho_\mu=\frac{\CK}{15}\p_{u^\mu}K\rightarrow \hat\rho_\mu u_\mu&=-\frac{2}{5}\trh t^3\,,	&	\rho_a&=\pm\frac{3\trh}{10}\CK_a\rightarrow t^2\trh^2=\mp\frac{5}{3}\Gamma\,,
\end{align}
equivalent to (4.25) in \cite{Camara:2005dc}.
\item Branch \textbf{A1}-\textbf{S2}
\begin{align}
\hat\rho_i=\frac{\CK}{30}\p_{u^i}K\rightarrow \hat\rho_0 u_0&=-\frac{4}{5}\trh t^3\, ,	&	\rho_a&=\pm\frac{\sqrt{6}\trh}{10}\CK_a\rightarrow t^2\trh^2=\mp\frac{5}{\sqrt{6}}\Gamma\,,
\end{align}
equivalent to (4.26) in \cite{Camara:2005dc}.
\item Branch \textbf{A2}-\textbf{S1}
\begin{align}
\hat\rho_\mu=\frac{\CK}{12}\p_{u^\mu}K\rightarrow \hat\rho_\mu u_\mu&=-\frac{1}{2}\trh t^3	\, ,&	\tilde\rho^a &=\pm\frac{\trh t^a}{2}\rightarrow t^2\trh^2=\frac{4}{3}\Gamma\,,
\end{align}
equivalent to (4.27)-(I) in \cite{Camara:2005dc}.
\item Branch \textbf{A2}-\textbf{S2}
\begin{align}
\hat\rho_i=\frac{\CK}{28}\p_{u^i}K\rightarrow \hat\rho_0 u_0&=-\frac{11}{14}\trh t^3\,,	&	\tilde\rho^a&=\pm\frac{\trh t^a}{14}\rightarrow t^2\trh^2=\frac{196}{99}\Gamma\,,
\end{align}
equivalent to (4.27)-(II) in \cite{Camara:2005dc}. 

\end{itemize}

\end{enumerate}

\section{Stability of the solutions} 
\label{s:stability}

Given the above families of extrema of the flux-induced potential, a natural question is which ones are actual vacua. In the following we would like to analyse this question at the classical level, by computing the spectrum of flux-induced masses on the former moduli fields. In particular, we will check whether the non-supersymmetric AdS extrema have any tachyonic direction with a mass below the BF found \cite{Breitenlohner:1982bm}. For simplicity, we will do this computation focusing only on the \textbf{A1-S1} and \textbf{A2-S1} branches, leaving the \textbf{S2} branch for further work.

\subsection{The Hessian}
\label{ss:hessian}

By construction, we have a potential whose first derivatives are of the form
\begin{equation}
\partial_\a V = \chi_\a^A \gamma_A \, ,
\end{equation}
with $\chi_a^A$ some regular functions of the saxions and the $\rho$'s. Therefore we have that
\be
\p_\a \p_\b V|_{\rm vac} = \chi_\a^A \p_\b \gamma_A \,,
\ee
where we have imposed our extremisation conditions $\gamma_A=0$. In fact, since $V_1$ is quadratic in the $\vec\gamma$, $\partial^2V_1|_{\rm vac}$ must be quadratic in $\partial \vec\gamma$. Indeed, one easily sees that
\begin{equation}
\boxed{ \p_\alpha\p_\beta V_1\rvert_{\rm vac}=2\left(\p_\alpha\vec{\g}^{\, t}\right) \hat{\bf Z}_{1} \left(\p_\beta\vec{\g}\right)}\
\label{HV1}
\end{equation}
where $\alpha=\{\xi^\mu,b^d,u^\delta,t^d\}$, $\hat{\bf Z}_{1}$ is defined as in \eqref{Zg1} and we have defined
\begin{align}
\vec{\g}^{\, t}=&\left(\begin{matrix}
\rho_0 &\g_a &\tg^a & \hg_\nu & \tr
\end{matrix}\right)\, , \\
\partial_{\xi^\mu}\vec{\g}^{\, t}=&\left(\begin{matrix}
h_\mu &0 &0 &0 & 0
\end{matrix}\right)\, ,\nonumber \\
\partial_{b^c}\vec{\g}^{\, t}=&\left(\begin{matrix}
\rho_c &\mk_{acd}\tr^d &\delta^a_c\tr &0 & 0
\end{matrix}\right)\, ,\nonumber \\
\partial_{u^\alpha}\vec{\g}^{\, t}=&\left(\begin{matrix}
0 &0 &0 &-\tr A\mk\p_\alpha\partial_\nu K-\trh\CK\p_\alpha\hat{\epsilon}^p_\nu & 0
\end{matrix}\right)\, ,\nonumber \\
\partial_{t^c}\vec{\g}^{\, t}=&\left(\begin{matrix}
0 & -2\tr C\mk_{ac} &-\tr B\delta^a_c &-3\tr A\mk_c\partial_\nu K & 0
\end{matrix}\right)\, . \nonumber
\end{align}
Notice that \eqref{HV1} is a product of two vectors with a positive definite metric. Therefore it corresponds to a positive definite Hessian, in agreement with the fact that $V_1$ is a sum of squares. 
The matrix of second derivatives of $V_2$ yields, by direct computation,
\begin{equation}
\boxed{\partial_\alpha \partial_\beta V_2\rvert_{\rm vac}=2\vec{\eta_\alpha}^t \hat{\bf Z}_{1} \p_\beta\vec{\gamma}= 2\p_\alpha\vec{\g}^{\, t} \hat{\bf Z}_{1} \vec{\eta_\beta}}\, 
\label{HV2}
\end{equation}
where we have defined
\begin{align}
\vec{\eta}_{\xi^\mu}^{\, t}=& \left(\begin{matrix}
0 &0  &0  &0 & 0
\end{matrix}   \right)\, ,
\\
\vec{\eta}_{b^d}^{\, t}=\tr&\left(\begin{matrix}
0 &0    &    \frac{3C}{\mk}K^{bc}\mk_{cd}    &    0 & 0
\end{matrix}\right)\, , \nonumber \\
\vec{\eta}_{u^\alpha}^{\, t}=&\tr\left(\begin{matrix}
0 & C\p_\alpha K\mk_a    &B\p_\alpha K t^a          &    \left(\frac{2}{3}-4A\right)\frac{\mk}{4} \left(\p_\alpha\p_\mu K-\p_\mu K \p_\alpha K\right)+e^{-K}\mk\tr K_{\beta\mu}\p_\alpha \left(e^KK^{\gamma\beta}\he^p_\gamma \right) & 0\end{matrix}\right)\, ,\nonumber \\
\vec{\eta}_{t^d}^{\, t}=&\tr\left(\begin{matrix}
0 & \frac{4C\mk}{3} K_{bd}    &    \frac{3B}{2\mk}K^{bc}\mk_{cd}     &    \mk\p_{t^d}\he^p_\mu & 0
\end{matrix}\right)\, .\nonumber
\end{align}
Unlike \eqref{HV1}, the term \eqref{HV2} is in general not definite, and may yield tachyonic directions. Putting both results together we find that the matrix of second derivatives is given by 
\begin{align}
\label{Hfin}
&\boxed{\p_\alpha\p_\beta V\rvert_{\rm vac}=2\left(\p_\alpha\vec{\g}^t\right) \hat{\bf Z}_{1} \left(\p_\beta\vec{\g}+\vec{\eta_\beta}\right)}\, 
\end{align}
which can also be written as:
\begin{align}
\p_\alpha\p_\beta V\rvert_{\rm vac}= \left(\p_\alpha\vec{\g_r}^t+\vec{\eta_\alpha}^t\right) \hat{\bf Z}_{1}\left(\p_\beta\vec{\gamma_r}+\vec{\eta_\beta}\right)+\left(\p_\alpha\vec{\g_r}^t\right) \hat{\bf Z}_{1} \left(\p_\beta\vec{\g_r}\right)-\vec{\eta_\alpha}^t \hat{\bf Z}_{1} \vec{\eta_\beta}\, .
\end{align}

\subsection{Flux-induced masses and perturbative stability}

The explicit form of the Hessian for the different branches {\bf A1-S1} and {\bf A2-S1} is given in Appendix \ref{ap:Hessian}, where the computation of its physical eigenvalues along  tachyonic directions is also performed. The relevant results for classical stability are summarised in table \ref{tablee}:

\begin{table}[H]
\begin{center}
\scalebox{1}{%
    \begin{tabular}{| c |     c | c | c | }
    \hline
 Branch & Tachyons & Physical eigenvalues & Massless modes \\
 \hline
  \textbf{A2-S1}& $0$    &    -& $2N$\\
 \hline
 \textbf{A1-S1, SUSY}&    $N$     &    $m^2_{tach}=\frac{8}{9}m_{BF}^2$&  $N$\\
  \hline
 \textbf{A1-S1, Non-SUSY}&     $N+1$    &    $m^2_{tach}=\frac{8}{9}m_{BF}^2$&  $N$\\
   \hline
    \end{tabular}}
    \caption{Massless and tachyonic modes for the extrema in the branch {\bf S1}. Here  $N$ stands for the number of complex structure moduli. The extra zero modes in the branch {\bf A2-S1} are discussed in appendix \ref{ap:complex}.}
    \label{tablee}
    \end{center}
\end{table}

Let us highlight some of the features resulting from this analysis:

\begin{itemize}

\item[-] Each vacuum has at least $N$ zero modes, which are the complex structure axions that do not appear in the superpotential \eqref{WQ}. As such, they do not appear in the F-term classical scalar potential, as one can check directly from eqs.\eqref{VF}-\eqref{ZAB}. Therefore they constitute $N$ flat directions of the classical potential. These unlifted axions may be eaten by D6-brane gauge bosons via the St\"uckelberg mechanism \cite{Camara:2005dc}.

\item[-] As expected from the analysis in \cite{Conlon:2006tq}, there are $N$ tachyons with mass $\frac{8}{9}|m_{BF}|^2$ in supersymmetric vacua. Such modes correspond to the saxionic directions that pair up with the flat axionic directions into complex fields. That is, they correspond to the saxions that do not appear in the superpotential \eqref{WQ}.

\item[-] As shown in appendix \ref{ap:ha1s1} the same tachyons are present in the non-supersymmetric vacua within the branch \textbf{A1-S1}, with the same mass in terms of the BF bound. Moreover, such non-SUSY vacua contain an extra tachyon which is a combination of complex and K\"ahler axionic directions, with exactly the same mass as the rest.  

\item[-] All these tachyons are absent in the \textbf{A2-S1} branch of solutions. Indeed, as shown in appendix \ref{ap:ha2s1}, all the solutions of this branch have a positive semidefinite Hessian. The tachyonic modes of the saxionic sector of the branch \textbf{A1-S1} are zero modes in this branch. They are however not flat directions and develop a positive quartic potential, see appendix \ref{ap:complex} for a detailed discussion. The rest of the spectrum does not arrange into mass-degenerate complex scalars. 

\item[-] A general analysis is more involved for the {\bf S2} branches. Following \cite{Camara:2005dc}, we have analysed them for the particular case of isotropic toroidal compactifications (i.e., where all three complex structure and three K\"ahler moduli are identified as $U_i = U$ and $T_i = T$, respectively). We have found that AdS solutions in this branch contain tachyons not satisfying the BF bound, and are therefore perturbatively unstable. It would be interesting to see if this feature is also present for more general solutions and compactifications within this branch.


\end{itemize}

\section{Validity of the solutions} 
\label{s:validity}

In the following we analyse the validity of our solutions from two different perspectives. On the one hand we will analyse the different scales from a 4d viewpoint. On the other hand we will comment on which 10d backgrounds could correspond to these 4d vacua.

\subsection{4d analysis and swampland conjectures}
\label{s:validity4d}

Since the different branches of solutions have been found via a classical potential $V$, one  should check that they fall in the compactification regime in which the corrections to $V$ are negligible. More precisely, a necessary condition to trust the above solutions is that the K\"ahler moduli are stabilised at sufficiently large values - so that $\alpha'$ corrections can be neglected - and the string coupling at small enough values - so that quantum corrections can also be neglected. In the following we will generalise the 4d validity analysis made in  \cite{DeWolfe:2005uu,Camara:2005dc} to our solutions, obtaining similar results. In short, the scaling of the volumes and couplings with the fluxes follows the same pattern as in these references, which allows to fall in the required regime for large values of the shifted four-form flux. Indeed, we have that
\begin{align}
&\ell_s^{-1} \hat{e}_a =  \tr \CK_a \left(C - \frac{1}{2} B^2 \right) \quad \rightarrow \quad t^2\left(C-\frac{B^2}{2}\right) \sim  \frac{\hat{e}}{m}\ ,
\\
&\hat{\rho}_\mu=\tr \mk A\p_\mu K \quad \rightarrow \quad u\sim \frac{t^3m A}{h}\, ,
\end{align}
where for simplicity we have assumed isotropic fluxes $h_\mu \sim h$, $\hat{e}_a \sim \hat{e}$. Since the $\hat{e}_a$ are unconstrained by tadpole equations, in principle we are free to scale them to be as large as needed. Assuming that $2C \neq B^2$, the moduli dependence on this scaling is given by
\begin{align}
&t\sim \hat{e}^{1/2}\, , &		 u&\sim \hat{e}^{3/2}\, .
\end{align}
In addition we have that
\begin{align}
&e^{-4D}\sim u^4 \sim \hat{e}^6 \rightarrow e^D\sim \hat{e}^{-3/2}\, , &	e^\phi&=\sqrt{{\rm Vol}_{\cM_6}}e^D\sim t^{-3/2}\sim \hat{e}^{-3/4}	\, .
\end{align}
These are the same scaling relations found in \cite{DeWolfe:2005uu} and so, for large $\hat{e}$, we are in a regime of large volume and weak coupling that prevents large corrections.  Finally, one can check that the four-form density scaling is similar to \cite{DeWolfe:2005uu} and therefore the corresponding higher derivative corrections are equally suppressed.

Additionally, one can check the scaling of the different mass scales, following for instance the relations given in \cite{Escobar:2018tiu}:
\begin{align}
\frac{M_{\rm KK}}{M_{\rm P}} & \sim \frac{g_s}{V^{2/3}}\sim t^{-7/2}\sim \hat{e}^{-7/4}\, , \nonumber\\
\frac{\Lambda}{M_{\rm P}^2} & \sim e^K\mk^2 m^2 \sim  \frac{t^3}{u^4}\sim t^{-3} \sim \hat{e}^{-9/2}\, , \\
R_{\rm AdS}M_{\rm P} & \sim \Lambda^{-1/2} M_{\rm P}\sim \hat{e}^{9/4}\, .\nonumber
\end{align}
We then recover the same scaling as found in \cite{DeWolfe:2005uu}, and in particular the same  parametric separation between the compactification scale and the AdS radius:
\begin{align}
\frac{R_{\rm AdS}}{R_{\rm KK}}&\sim \hat{e}^{1/2}\, ,	& \hat{R}_{\rm KK}\sim \hat{R}_{\rm AdS}^{7/9}\ \longrightarrow	\ \frac{{M}_{\rm KK}}{M_{\rm P}} \sim \hat{\Lambda}^{7/18}\, ,
\end{align}
where $\hat{R} = R M_{\rm P}$ and $\hat{\Lambda} = \Lambda/M_{\rm P}^2$. Lastly, using the results derived in appendix \ref{ap:ha1s1}, for the \textbf{A1-S1} branch we have:
\begin{equation}
\label{amigo}
m^2_{\text{moduli}}\sim m_{BF}^2\sim \Lambda\sim\frac{1}{R_{AdS}^2}\, ,
\end{equation}
where $m^2_{\text{moduli}}$ refers to the canonically normalised mass of the moduli becoming massive. 

Regarding the swampland conjectures, the last relation \eqref{amigo} satisfies the criterium suggested in  \cite{Gautason:2018gln} for the lightest scalars. It would be interesting to check if the spectrum of  St\"uckelberg masses associated with the zero modes as well as the spectrum of the other branches still satisfy this relation. In terms of the recent AdS conjectures formulated in \cite{Lust:2019zwm}, all the AdS vacua found in our analysis satisfy the plain AdS Distance Conjecture, while the supersymmetric ones would fail to satisfy its strong version. It was suggested in \cite{Lust:2019zwm} that this failure could be related to the lack of knowledge of the full 10d supergravity background describing such vacua. In fact, to date the absence of a solution to the 10d equations of motion holds for each of the AdS vacua found in our 4d analysis, and is to be expected that finding their 10d description is at the same level of difficulty. In the following we will comment on certain characteristics that such a 10d solution should have.

\subsection{Towards a 10d description}

Massive type IIA 10d supergravity solutions of the form AdS$_4 \times X_6$ are relatively well understood in several instances, like when the internal manifold $X_6$ is endowed with a SU(3)-structure underlying 4d $\CN=1$ supersymmetry \cite{Behrndt:2004km,Behrndt:2004mj,Lust:2004ig}. In that case, the 10d background supersymmetry equations can be written as \cite{Grana:2005sn,Koerber:2007jb}
\bes
\label{adssu3}
\begin{equation}
\label{gauge}
d_{H} \RE \Omega \, =\, e^\phi *_6 \left(G_0 - G_2 + G_4 - G_6 \right) + 3 \RE (w_0e^{iJ})\, ,
\end{equation}
\begin{equation}
\label{DW}
d_{H} e^{iJ} \, =\, 2i \bar{w}_0 \IM \Omega\ ,
\end{equation}
\begin{equation}
\label{string}
d_{H}  \IM \Omega \, =\, 0\, ,
\end{equation}
\ees
where $\phi$ is constant, $d_{H} = d + H \wedge$ and $w_0 = e^{i\th}/R\in \mathbb{C}$ is the constant entering the Killing equation of an AdS$_4$ of radius $R\ell_s$. For $w_0 \in i \mathbb{R}$ the solution to the above equations can be parametrised as
\be\label{solutionsu3}
G_6= 0\, , \quad \quad H = \frac{2}{5} e^{\phi} G_0 \,  \IM \Omega\, , \quad \quad G_2 = e^{-\phi} W_2 \, , \quad \quad G_4 = \frac{3}{10} G_0 J \wedge J \, ,
\ee
where $G_0 =  5e^{-\phi}\, \IM w_0$ is a constant and $W_2$ is a real primitive $(1,1)$-form, namely a SU(3) torsion class of $\Omega$ \cite{Chiossi:2002tw}. Notice that because of \eqref{gauge} $W_2$ cannot have a harmonic component. One may now express the RR background fluxes as
\be\label{expflux}
G_0 = -\tilde{\rho}\, ,  \quad G_2 = \tilde{\rho}^a\, \ell_s^{-2} \omega_a + \alpha_2 \, ,  \quad G_4 = - \rho_a\, \ell_s^{-4}  \tilde{\omega}^a  +\alpha_4 \, , \quad G_6 = \rho_0\, \ell_s^{-6} \frac{d{\rm vol}_{X_6}}{{\rm vol}_{X_6}} + \alpha_6\,, 
\ee
where $\omega_a$ and $\tilde{\omega}^a$ are a basis of harmonic two- and four-forms of $X_6$, and $\alpha_{2p}$ are globally well-defined forms with no harmonic component. We also expand the NS-flux as in \eqref{Hflux}. One can then see that \eqref{solutionsu3} amounts to apply the Ansatz \eqref{solutions} with the choice of constants $A, B, C$ corresponding to the supersymmetric {\bf A1-S1} branch, together with $\alpha_2 = e^{-\phi} W_2$ and $\alpha_4 = \alpha_6 =0$. Even if a $W_2 \neq 0$ signals that the metric on $X_6$ is not Calabi-Yau, supersymmetry requires that $W_2$ has no harmonic component, just as in type IIA compactifications to 4d Minkowski  \cite{Marchesano:2014iea}. As such, its presence can be considered as a deformation of the Calabi-Yau metric similar to a warp factor, rather than a discrete deformation or genuine geometric flux carrying topological information, see e.g. \cite{Tomasiello:2005bp,Marchesano:2006ns}. 

Despite these suggestive features, one can see that the \eqref{adssu3} is too simple to describe an actual 10d background corresponding to a type IIA compactification with fluxes, O6-planes and D6-branes. First, it features a constant dilaton and warp factor, which are in tension with the backreaction of such localised sources. Second, it implies that $\Omega \wedge G_2 \equiv 0$, which is never true in the vicinity of a D6-brane or O6-plane. Finally, it is incompatible with the Bianchi identity for $G_2$. This reads
\be\label{BIG2}
d G_2 + HG_0 = -\sum_a \delta(\Pi_a)\quad \rightarrow \quad dW_2 = - e^{\phi} \left[G_0 H + \sum_a \delta(\Pi_a)\right]\, ,
\ee
where $\delta(\Pi_a)$ are bump functions localised on the three-cycles $\Pi_a$ wrapped by the D6-branes and O6-planes, and include their relative charge. As usual, RR tadpole cancellation amounts to require that the quantity in brackets vanishes in cohomology, so that $G_2 = e^{-\phi}W_2$ can be globally well-defined. However, \eqref{BIG2} together with $\Omega \wedge G_2 \equiv 0$ implies a flux density $|G_4|^2$ which is negative in the bulk and singular  on top of any localised source. A proposal to circumvent these problems is to modify the Bianchi identity \eqref{BIG2} by replacing the localised sources with smeared ones \cite{Acharya:2006ne}, so that one can take $G_2 \equiv 0$. 

Instead of modifying the Bianchi identity, one may try embed the above solution into a type IIA AdS$_4$ compactification based on a background with $SU(3) \times SU(3)$ structure, which is compatible with non-trivial dilaton and warp factor \cite{Lust:2008zd,Lust:2009zb,Lust:2009mb}. In the language of \cite{Grana:2005sn,Koerber:2007jb} in this more general case the supersymmetry equations \eqref{adssu3} are replaced by
\bes
\label{adssu3su3}
\begin{equation}
\label{gauge33}
d_{H} \left( e^{4A-\phi} \RE \Psi_1\right) \, =\, e^{4A} *_6 \left(G_0 - G_2 + G_4 - G_6 \right) + 3 e^{3A-\phi} \RE (w_0 \Psi_2)\, ,
\end{equation}
\begin{equation}
\label{DW33}
d_{H} \left[\IM \left(w_0 e^{3A-\phi} \Psi_2 \right)\right] \, =\, 2 |w_0|^2 e^{2A-\phi} \IM \Psi_1\ ,
\end{equation}
\begin{equation}
\label{string33}
d_{H} \left( e^{2A-\phi}  \IM \Psi_1 \right)\, =\, 0\, ,
\end{equation}
\ees
where $\Psi_1$ and $\Psi_2$ are odd and even polyforms, respectively, replacing $\Omega$ and $e^{iJ}$. Again, one may plug the expression for the fluxes \eqref{expflux} and solve for the above equations. The polyforms $\Psi_1$, $\Psi_2$ will have a more involved profile than in their $SU(3)$-structure counterparts, but if $w_0 \in i \mathbb{R}$ and their harmonic components are the same as before, so will be the harmonic components of the background fluxes. That is, one would be recovering a flux background whose projection into harmonic forms reproduces the 4d supersymmetric solution of the {\bf A1-S1} branch. In that case, because the rest of the background would be encoded in non-harmonic $p$-forms, it could be considered as a deformation of the naive Calabi-Yau geometry, which one would expect to asymptotically recover in the dilute flux limit. Thus, it would be interesting to see if the background \eqref{adssu3su3} with these characteristics can be made compatible with the actual Bianchi identity \eqref{BIG2}. If that is the case, one may also attempt to provide the 10d description of the non-supersymmetric AdS vacua of section \ref{s:IIAvacua}, perhaps by identifying them with the AdS backgrounds in \cite{Lust:2008zd}.

\section{Including mobile D6-branes}
\label{s:D6branes}

As in \cite{Grimm:2011dx,Kerstan:2011dy,Carta:2016ynn} we may generalise the above setup by considering type IIA orientifold compactifications where D6-branes have deformation and Wilson line moduli. In order to preserve supersymmetry such D6-branes must wrap special Lagrangian three-cycles $\Pi_\alpha \subset {\cal M}_6$ with vanishing worldvolume flux \cite{Becker:1995kb,Martucci:2005ht}. Then the open string moduli space is characterised by $b_1(\Pi_\a)$ complex moduli \cite{Mclean96deformationsof,Hitchin:1997ti}. These are defined as \cite{Herraez:2018vae,Escobar:2018tiu}
\begin{equation}
\Phi_\alpha^i = T^a f_{\alpha\, a}^i - \theta^i_\alpha = \hat  \theta_\alpha^i + i\, \phi^i_\alpha\,  ,
\end{equation}
where $i$ runs over the integer harmonic one-forms $\zeta_i$ of $\Pi_\alpha$, $\theta^i_\alpha$ is the Wilson line corresponding to each of them and $f_{\alpha\, a}^i$ is a function of the corresponding geometric deformation of $\Pi_\alpha$ defined in terms of a chain integral. We refer the reader to \cite{Carta:2016ynn,Herraez:2018vae,Escobar:2018tiu} for further details on these definitions. 

For each harmonic one-form $\zeta_i \in H^1(\Pi_\a, \mathbb{Z})$ there is  a two-form $\eta^i \in H^2(\Pi_\a, \mathbb{Z})$ along with a worldvolume flux $F = n_{F\, i}^\alpha$ that can be turned on. This enters the D6-brane DBI action and therefore the scalar potential in the combination $n_{F\, i}^\a - \oh g_{i\, \a}^\mu h_\mu$, where $g_{i\, \a}^\mu$ is also defined in terms of a chain integral \cite{Carta:2016ynn,Herraez:2018vae,Escobar:2018tiu}. As such the presence of such fluxes generates a potential, captured by the superpotential
\be
\ell_s {W}_{\rm D6}  \, = \, \Phi_\a^i( n_{F\, i}^\a - n_{a\, i}^\a T^a ) + \ell_s W_0\, ,
\label{WD6}
\ee
where 
\be
n_{a\, i}^\a  =  \frac{1}{\ell_s^3} \int_{\Pi_\a} \omega_a \wedge \zeta_i \in \mathbb{Z}
\label{nai}
\ee
are non-vanishing whenever the two-cycles of $\Pi_\alpha$ are non-trivial in $H_2(\cM_6, \mathbb{Z})$. Indeed, as pointed out in \cite{Marchesano:2014iea} in this case the open string moduli develop a potential due to the D6-brane backreaction on a compact space.

An important effect to take into account is the field redefinition of the closed string moduli in the dilaton-complex structure sector in the presence of open string moduli. We have that the new variables read \cite{Herraez:2018vae,Escobar:2018tiu}
\be
U^\mu = U_\star^\mu  + \oh \sum_\a \left(g^\mu_{i\, \a} \th^i_\a -T^a  H_{a \, \a}^\mu \right)  \, ,
\label{redefN}
\ee
where $U_\star^\mu = (N^K_\star, U_{\Lambda\, \star})$ stand for the complex structure moduli in the absence of mobile D6-branes, namely \eqref{cpxmoduli}, and $U^\mu = (N^K, U_\Lambda)$ are the redefined 4d variables. Finally $H_{a \, \a}^\mu$ are functions of the saxions defined in terms of $f_{\alpha\, a}^i$ and $g_{i\, \a}^\mu$ \cite{Herraez:2018vae}. Notice that \eqref{KQ} is a function of $u^\mu_\star$, which is to be written in terms of the new 4d variables by means of \eqref{redefN}. 

A similar statement holds for the scalar potential, which still has the form \eqref{VF} but now with
\begin{equation}
{\bf Z} = 
\left(
\begin{matrix}
 4 & 0 & 0 & 0 & 0 & 0 & 0 \\
 \\
 0 & K^{ab} & 0 & 0 & 0 & 0 & 0 \\
 \\
 0 & 0 & \frac{4}{9} \mk^2 K_{ab} & 0 & 0 & 0 & 0 \\
 \\
  0 & 0 & 0 & G^{ij} & 0 & 0 & 0\\
 \\
  0 & 0 & 0 & 0 & t^{a}t^b G^{ij} & 0 & 0\\
  \\
0 & 0 & 0 & 0 & 0 &  K^{\mu \nu }  & \frac{2}{3} \mk u^{\mu}_\star  \\
 \\
 0 & 0 & 0 & 0 & 0 & \frac{2}{3}\mk  u^{\nu}_\star  &\frac{\mk^2}{9} \\
\end{matrix}
\right)\, , \qquad 
\vec{\rho} \, =\, 
\left(
\begin{array}{c}
\rho_0' \\ \rho_a' \\ \tilde{\rho}^{\prime \, a} \\ \rho_i' \\ \rho_{a\, i} \\ \hat\rho_\mu  \\ \tilde{\rho}
\end{array}
\right)\, ,
\label{ZrhoD6}
\end{equation}
where
\begin{align}
 \r_0'&=\r_0+\hth^i\r_i\, ,\nonumber
\\
\r_a'&=\r_a-\hth^i\r_{ai}+f^i_a\r_i-\frac{1}{2}H^\mu_a\hr_\mu\, ,
\nonumber
\\
 \tr'^a&=\tr^a-\left(\mk^{ab}\ph^i+\mk^{ac}t^bf^i_c\right)\r_{bi} \label{newrhos}\, ,
\\
\r_i'&=\underbrace{\ell_s^{-1} \left(n_{i}-b^an_{ai}\right)}_{\r_i} -\frac{1}{2}g^\mu_{i}\hr_\mu=\r_i -\frac{1}{2}g^\mu_{i}\hr_\mu\, ,
\nonumber
\\
\r_{ai}&= \ell_s^{-1} n_{ai}\, . \nonumber
\end{align}
A few comments are in order. To simplify the notation we have absorbed the D6-brane index $\alpha$ into the open string moduli index $i$. Here $\rho_0$ is defined as in \eqref{rhos} but now in terms of the redefined RR axion $\xi^\mu = \xi_\star^\mu - \frac{1}{2} b^a H_a^\mu + \frac{1}{2} g_i^\mu \theta^i$. Finally, notice that the $\rho$'s defined in \eqref{newrhos} not only depend on fluxes and axions, but also on some saxions,  differently from those defined in \cite{Herraez:2018vae}. This is just as well for the purpose of this analysis, as we are going to combine them right away in terms of saxion-dependent  polynomials $\gamma_A$. Indeed, applying the strategy of section \ref{ss:ansatz} we define
\begin{equation}
\label{gammaopen}
\vec{\gamma}'=
\left(\begin{matrix}
\g'_0\\
\g'_a\\
\tg'^a\\
\g'_i\\
t^a\rho_{ai}\\
\hg_\mu\\
\tr
\end{matrix}\right)=\left(\begin{matrix}
\r'_0\\
\r'_a-\tr\e_a\\
\tr'^a-\tr\te^a\\
\r'_i\\
t^a\rho_{ai}\\
\hr_\mu-\tr\he_\mu\\
\tr
\end{matrix}\right)\, ,
\end{equation}
where we are not relabelling $t^a\rho_{ai}$ in order to not to overload the notation and, as before, we assume that each of the terms of this vector vanishes in the vacuum, except $\tr$. The potential can again be split in two terms
\begin{align}
\label{openpotential}
V'&=\underbrace{e^K\left[\frac{4}{9}\mathcal{K}^2\tilde{\gamma}'^a\tilde{\gamma}'^b K_{ab}+4\g_0'^2+K^{ab}\gamma_a'\gamma_b'+K^{\mu\nu}\hat{\gamma}_\mu\hat{\gamma}_\nu+G^{ij}\left(\g'_i\g'_j+t^a\r_{ai}t^b\r_{bj}\right)\right]}_{V_1'}+\nonumber\\&+\underbrace{e^K \left[\frac{4}{3}\mathcal{K}\tilde{\rho}u_\star^\mu\hat{\gamma}_\mu+\frac{8}{9}\mathcal{K}^2\tilde{\rho}K_{ba}\tilde{\gamma}'^a\tilde{\epsilon}^b+2\tilde{\rho }K^{ab} \epsilon _a \gamma'_b+2\tilde{\rho} K^{\mu \nu }\gamma _{\mu } \hat{\epsilon }_{\nu }+\tilde{\rho}^2\left(\alpha+\frac{\mathcal{K}^2}{9}\right)\right]}_{V_2'}\, ,
\end{align}
where $\alpha=K^{ab}\epsilon_a\epsilon_b+\frac{4}{9}\mathcal{K}^2K_{ba}\tilde{\epsilon}^b\tilde{\epsilon}^a+K^{\mu \nu }\hat{\epsilon }_{\nu } \hat{\epsilon }_{\nu }+\frac{4}{3} \mathcal{K}u_\star^{\mu } \hat{\epsilon }_{\nu }$ and
\begin{align}
\te^a&=Bt^a\, ,	&	&\e_a=C\mk_a\, ,	&	\he_\mu=A\mk\p_\mu K+\mk\te^p_\nu\, .
\end{align}

Again, as $V_1$ is quadratic in the $\gamma'_A = \{\gamma'_0, \gamma'_a, \tilde{\gamma}'^a, \gamma'_i, t^a\rho_{ai}, \hat{\gamma}_\mu \}$, the extremisation conditions only depend on $V_2$. As before, one can take derivatives of $V_2$ along axionic and saxionic directions, and impose an Ansatz of the form \eqref{Ansatz}. The discussion parallels to a large extent the one in section \ref{ss:branches}, so we will provide fewer details of the derivation.

\subsubsection*{Axionic sector}
\begin{align}\nonumber
\p_{ \xi^\mu} V_2' & =8e^K\g_0'\hat\rho_\mu\,, \\
\p_{\hth^i} V_2' & =-\frac{8\mk e^K}{3}C\tr \left( t^b \rho_{b\, i}\right)\,, \\ \nonumber
\p_{b^a} V_2' & = e^K\frac{2}{3}\CK \tr \left[ \CK_{a} B \tilde{\rho} + 4 C t^b \left( \CK_{abc}\tilde{\rho}'^c + f_a^i \rho_{b\, i}\right)\right] 
 =  e^K\frac{8}{3}\CK \tr   \CK_{ac} \left( C \tilde{\gamma}'^c + B \left( C+\frac{1}{4}\right) t^c \tilde{\rho}\right) \,. 
\end{align}
The last expression is linear on $\tilde{\gamma}'^a$ for either $B=0$ (branch {\bf A1}) or $C=-\frac{1}{4}$ (branch {\bf A2}).

\subsubsection*{Saxionic sector}
\begin{align}
\partial_{u^\alpha}V_2'&=\left(\frac{1}{3}-2A\right)4\mk e^K\tr\left(u^\mu\p_\alpha K+\delta^\mu_\alpha\right)\hat{\gamma}_\mu+\p_\alpha K \left(\frac{2Be^K}{3}\mathcal{K}\tilde{\rho}\mk_b\right)\tilde{\gamma}'^b \\ &+\p_\alpha K\left(\frac{8C}{3} e^K\tilde{\rho }\mk t^b\right) \gamma'_b+2\mk\tr\p_\alpha\left(e^KK^{\mu\nu}\te^p_\nu\right)\hg_\mu\nonumber\\& +\trh^2e^{K}\left[\p_{u^\sig} K \left( \frac{\CK^2}{9} + \a\right) + \left( \p_{u^\sig} K^{\mu\nu}\right) \hat\eps_\mu \hat\eps_\nu + \frac{4}{3} \CK  \hat\eps_\sig \right]\, ,\nonumber\\
\partial_{t^a}V_2'&=\frac{\left(H^\alpha_a-f^i_ag^\alpha_i\right)\p_{u^\alpha} V_2}{2}+\left(\frac{4Be^K}{3}\mathcal{K}\tilde{\rho}\mk_{ab}\right)\tilde{\gamma}'^b+\left(\frac{8C}{3} e^K\tilde{\rho }\mk \right)\gamma'_a \\ &+\left(2e^K\tr\mk K^{\mu\nu}\p_{t^a}\te^p_\mu\right)\hat{\gamma}_\mu-\frac{4e^K}{3}\mk\tr f^i_a\left(2C\r'_i-Bt^d\r_{di}\right)\nonumber\\&+e^K\trh^2 \left[\p_{t^a} K \left( \frac{\CK^2}{9} + \a\right) +\frac{1}{9} \p_{t^a} \CK^2 + \left( \p_{t^a} K^{bc}\right) \eps_b\eps_c + \frac{4}{9}\p_{t^a} \left( \CK^2K_{ab}\right) \tilde{\eps}^a\tilde{\eps}^b + {4} \CK_a  u^\mu \hat\eps_\mu\right]\,, \nonumber \\
\partial_{\phi^i}V_2'& =\frac{g^\mu_i\p_{u^\mu} V_2}{2}+\frac{4e^K}{3}\mk\tr\left(2C\r_i'-Bt^b\r_{bi}\right)\,.
\end{align}
One can see that the conditions for $\p V$ to be linear on the $\gamma_A'$ are exactly the same as in the case without mobile D6-branes with the extra conditions $\{\gamma_i'=\r'_i=0, t^a\rho_{ai}=0\}$. Therefore, the same branches of vacua are recovered replacing the previous $\gamma$'s by the new ones. Notice that one of these branches corresponds to non-supersymmetric Minkowski vacua with D6-branes and that the conditions \eqref{Ansatz} precisely reproduce those of the vacua found in \cite{Escobar:2018tiu}. In general, we expect that the vacua of section \ref{s:stability} remain perturbatively stable in the presence of mobile D6-branes, generalising the results of \cite{Escobar:2018tiu} to AdS vacua. A detailed analysis of the Hessian, whose expression is given in appendix \ref{ap:openH}, is however left for future work.


\section{Conclusions}
\label{s:conclu}

In this paper we have performed a general search for vacua of the classical type IIA flux potential in generic Calabi-Yau orientifold compactifications. Our analysis extends the one made in \cite{DeWolfe:2005uu} in the sense that we allow for non-supersymmetric vacua as well, the only requirement being the Ansatz of section \ref{ss:ansatz}. Implementing it we find several branches of vacua, including the supersymmetric AdS branch of \cite{DeWolfe:2005uu}, a Minkowski $\CN=0$ branch mirror to type IIB with three-form fluxes \cite{Palti:2008mg} and several new branches of non-supersymmetric AdS vacua. Remarkably, when restricted to the isotropic torus, these branches reduce to precisely the ones found in \cite{Camara:2005dc}. In this sense, our results can also be seen as an extension of the AdS type IIA flux landscape familiar from toroidal compactifications to the plethora of Calabi-Yau geometries. 

The technical ingredient behind this progress is essentially the bilinear form of the flux potential developed in \cite{Bielleman:2015ina,Carta:2016ynn,Herraez:2018vae}. This expression for $V$ conveniently factorises the saxionic and axionic degrees of freedom of the compactification, and arranges the latter in flux-axion polynomials $\rho_A$ invariant under discrete shift symmetries. This permits a more economic and organised description of the extrema conditions and their solutions, which arrange themselves into branches parametrised by real constants $A, B, C$ - see table \ref{vacuresul}. Moreover, it also allows incorporating into the analysis the light degrees of freedom of mobile D6-branes, together with their worldvolume fluxes. As a result one is able to extend the above landscape of solutions to the open string sector, in the spirit of \cite{Gomis:2005wc}.

Given these branches of critical points of the potential, the next step is to verify if they correspond to (possibly metastable) vacua. We have performed the analysis of the classical stability for the simplest branches of solutions, namely the {\bf Branch S1} of section \ref{ss:branches}, where the homogeneity properties of the K\"ahler potential allow to compute the mass spectrum of the would-be moduli. We have compared such masses with the Breitenlohner-Freedman bound, finding that {\it i)} the {\bf Branch A1-S1} develops tachyons satisfying the bound and {\it ii)} the {\bf Branch A2-S1} is absent of any tachyons. Therefore, this set of extrema already constitute a Landscape of AdS flux vacua. It would remain to analyse the non-perturbative stability of this collection of vacua, which could represent an interesting playground to test the recent conjecture \cite{Ooguri:2016pdq} on $\CN=0$ AdS compactifications.

Our results can be applied and generalised in different directions. For instance, they could be extended to include non-Calabi-Yau geometries, like SU(3) compactifications with metric fluxes. Indeed, such compactifications can also be described by an effective scalar potential bilinear in the fluxes \cite{House:2005yc,Camara:2005dc,Caviezel:2008ik} so in principle our strategy should apply to them as well. In fact, such bilinear structure arises as well in any supersymmetric effective field theory based on three-forms, like the ones recently developed in \cite{Farakos:2017jme,Bandos:2018gjp,Lanza:2019xxg}. One could combine our results with the said formalism to have a (partial) EFT description of the landscape of AdS flux vacua, together with membrane-mediated transitions between them. In this context, one may analyse the phenomenological properties of this landscape of vacua as an ensemble \cite{Douglas:2003um}.  For instance, given the F-terms for each of these vacua, one could extend the analysis of \cite{Escobar:2018tiu} to compute the spectrum of supersymmetry-breaking soft terms induced on the open string sector, and then analyse its statistical distribution. 

For each of these developments, a crucial step is to establish the perturbative stability of the extrema of the potential. In this sense, it would be interesting to extend the results of Appendix B to other branches not analysed in there, including solutions with mobile D6-branes. The same type of analysis could also be carried out for further examples of classical AdS vacua, like those involving metric fluxes. Some of these have the advantage that their 10d description is well understood, so analysing them with the formalism used here for Calabi-Yau orientifolds may help to better understand the 10d description of the latter. In general, we expect that a global understanding of type IIA flux vacua from a 4d perspective will shed light on their microscopic description, helping to comprehend the ensemble of type IIA flux compactifications and eventually the string Landscape.

\bigskip

\vspace*{.05cm}

\centerline{\bf  Acknowledgments}

\vspace*{.05cm}

\bigskip
We would like to thank \'Alvaro Herr\'aez, Luis E. Ib\'a\~nez and Eran Palti for useful discussions.  This work is supported by the Spanish Research Agency (Agencia Estatal de Investigaci\'on) through the grant IFT Centro de Excelencia Severo Ochoa SEV-2016-0597, and by the grants FPA2015-65480-P and PGC2018-095976-B-C21 from MCIU/AEI/FEDER, UE. J.Q. is supported through the FPU grant No. FPU17/04293.
F.M. gratefully acknowledges support from the Simons Center for Geometry and Physics, Stony Brook University and from a grant from the Simons Foundation at the Aspen Center for Physics. Some of the research for this paper was performed at both institutions.



\appendix


\section{Some useful relations}
\label{ap:relations}
As reviewed  in section \ref{s:IIAorientifold}, type IIA compactifications on Calabi-Yau orientifolds come with moduli spaces parameterised by K\"ahler moduli and complex structure moduli. These moduli spaces are endowed with a K\"ahler geometry with the K\"ahler metric being proportional to the second derivative of the  K\"ahler potential:
\begin{align}
K=K_T+K_Q&=-\log (\mathcal{G}_T\mathcal{G}_Q^2)=	-\log (\mathcal{G})	&	 &\longrightarrow 	&	&K_{AB}=\frac{1}{4}\p_A\p_B K\, ,
\end{align}
where, following our notation, $A=\{t^a,u^\alpha\}$, $\mathcal{G}_T=\frac{4}{3}\mk$ is a homogeneous function of degree three on the $t^a$ and $\mathcal{G}_Q^2$ is a homogeneous function of degree four on the $u^\alpha$. These properties, along with the fact that $K_{a\mu}=0$, allow to compute some useful relations regarding the K\"ahler potential and the  K\"ahler metric. Let us start by noting that:
\begin{align}
t^a\p_{t^a} &\mathcal{G}=3\mathcal{G}	\, ,&	u^\mu\p_{u^\mu} &\mathcal{G}=4\mathcal{G}\, .
\end{align}
It will be  convenient to write the explicit form of the metric in each sector:
\begin{align}
&K_{ab}=\frac{3}{2\mathcal{K}}\left(\frac{3\mathcal{K}_a\mathcal{K}_b}{2\mathcal{K}}-\mathcal{K}_{ab}\right)	\, ,&	
&K_{\mu\nu}=\frac{1}{4}\left(\frac{\p_\mu G\p_\nu G}{G^2}-\frac{\p_\mu\p_\nu G}{G}\right)\, ,
\end{align}
with $\mk_{ab}=\mk_{abc}t^c$, $\mk_{a}=\mk_{abc}t^bt^c$, and $G= {\cal G}_Q^2$. Then, it is  straightforward to check that:
\begin{multicols}{2}
\begin{itemize}
\item $u^\mu \partial_\mu K=-4$,
\item $K^{\mu\nu}\partial_\mu K=-4u^\nu$,
\item $K^{\mu\nu}\partial_\mu K\partial_\nu K=16$,
\item $\p_\alpha K^{\mu\nu}\partial_\mu K=-8\delta^\nu_\alpha$,
\item $K^{\mu\nu}\p_\alpha\p_\mu K\p_\nu K=4\p_\alpha K$,
\item $\p_\alpha\p_\beta K^{\mu\nu}\p_\mu K\p_\nu K=8\p_\alpha\p_\beta K$,
\item $u^\alpha\partial_\alpha K_{\mu\nu}=-2K_{\mu\nu}$,
\item $u^\alpha\partial_\alpha K^{\mu\nu}=2K^{\mu\nu}$,
\item $u^\mu\p_\mu\p_\alpha K=-\p_\alpha K$;
\end{itemize}
\end{multicols}
and:
\begin{multicols}{2}
\begin{itemize}
\item $K^{ab}\mathcal{K}_a\mathcal{K}_b=\frac{4}{3}\mathcal{K}^2$,
\item $\mathcal{K}^{cd}\mathcal{K}_d=t^c$,
\item $K^{cd}\mathcal{K}_d=\frac{4}{3}\mathcal{K}t^c$,
\item $\mathcal{K}^{bd}K_{ab}=-\frac{1}{4\mathcal{K}}\left(6\delta^d_a-\frac{9t^d\mathcal{K}_a}{\mathcal{K}}\right)$,
\item $t^cK_{ca}=\frac{3}{4}\frac{\mathcal{K}_a}{\mathcal{K}}$,
\item $K^{cd}\mathcal{K}_{da}=\mathcal{K}\left(-\frac{2}{3}\delta^c_a+\frac{2t^c\mathcal{K}_a}{\mathcal{K}}\right)$,
\item $t^a\partial_aK_{bc}=-2K_{bc}$,
\item $t^a\partial_aK^{bc}=2K^{bc}$,
\item $\p_a\mk^{cb}\mk_c=-\delta^b_a$,
\item $\p_a K^{bc}\mk_b=\frac{8}{3}\mk \delta^c_a$;
\end{itemize}
\end{multicols}
\noindent
where implicitly we have defined $\p_\alpha\equiv \p_{u^\alpha}$,  $\p_a\equiv \p_{t^a}$,  $K^{\mu\nu}K_{\mu\eta}=\delta^\nu_\eta$, $K^{ab}K_{bc}=\delta^a_c$ and $\mk^{ab}\mk_{bc}=\delta^a_c$.

\section{Analysis of the Hessian}
\label{ap:Hessian}

In this appendix we analyse the properties of the matrix of second derivatives of the potential, or Hessian. As discussed in section \ref{ss:hessian}, due to our Ansatz \eqref{Ansatz} the Hessian can be written as
\begin{align}
H_{\a\b}&=\p_\alpha\p_\beta V\rvert_{\rm vac}=2\left(\p_\alpha\vec{\g}^t\right) \hat{\bf Z}_{1} \left(\p_\beta\vec{\g}+\vec{\eta_\beta}\right)\,.
\label{ap:Hfin}
\end{align}
For the solutions in table \ref{vacuresul} within the branches {\bf A1-S1} and {\bf A2-S1} (or equivalently for the solutions of the form \eqref{solutions} with $\hat{\epsilon}^p_\mu=0$) one can write and explicit expression for {\bf H} in terms of the parameters $A,B, C \in \mathbb{R}$. Ordering the derivatives as $\left(\p_{\xi^\mu},\p_{b^a},\p_{u^\alpha}, \p_{t^a} \right)$ one finds that:
\begin{align}
\label{gZ1g}
&\left(\p_\alpha\vec{\g_r}^t\right) \hat{\bf Z}_1 \left(\p_\beta\vec{\g_r}\right)=\nonumber\\&e^K\left(\begin{matrix}
4A^2\mk^2\tr^2\p_\nu K\p_\mu K & 4 AC\mk\tr^2\partial_\nu K\mk_a	&0 &	0\\
\\
4 AC\mk\tr^2\partial_\mu K\mk_b&-C_1^2\mk\mk_{ab}+C_2^2\mk_a\mk_b &0 &	B\left(C_3\mk \mk_{ab} -C_4\mk_a\mk_b\right)\\
\\
0 &0 &C_5^2\mk^2K_{\alpha\beta}  &	\frac{3C_5^2}{4}\mk\partial_\alpha K\mk_b\\ 
\\
 0 &B\left(C_3\mk \mk_{ab} -C_4\mk_a\mk_b\right) &\frac{3C_5^2}{4}\mk\partial_\alpha K\mk_a&	-C_6^2\mk\mk_{ab}+C_7^2\mk_a\mk_b \\\\
\end{matrix}\right)\,,
\end{align}
with 
\begin{align}
C_1^2&=\frac{2\tr^2}{3}\left(1+B^2\right),\nonumber	&	C_2^2&=\tr^2\left(1+2B^2+4C^2\right),\nonumber\\
C_3&=\frac{2}{3}\tr^2\left(1+2C\right),	&	C_4&=\tr^2\left(1+4C\right),\nonumber\\
C_5^2&=16A^2\tr^2,\nonumber & C_6^2&=\frac{2}{3}\tr^2\left(B^2+4C^2\right),\nonumber\\
C_7^2&=\tr^2\left(B^2+8C^2+144A^2\right);
\end{align}
and that
\begin{align}
\label{gZ1eta}
\left(\p_\alpha\vec{\g_r}^t\right) \hat{\bf Z}_1\vec{\eta_\beta}&=e^K\left(\begin{matrix}
0&0&0&0\\
0 &  \sigma_1\mk_{ab}  &  0  &     -B\sigma_1\mk_{ab}\\
  \\
0 & 0	&	A\sigma_2\mk\left(\p_\alpha K	\p_\beta K-4 K_{\alpha\beta}\right)	&	0\\
  \\
 0 &  -B\sigma_1\mk_{ab}		&	0	& -\left(B\s_3+2C\s_1\right)\mk_{ab}\\
 \end{matrix}\right)\, ,
\end{align}
with $\sigma_1=\frac{4C\mk}{3}\tilde{\rho}^2$,  $\sigma_2=\left(\frac{1}{3}-2A\right)2\mk \tr^2$ and  $\sigma_3=\frac{2B}{3}\mathcal{K}\tilde{\rho}^2$.

Already from this expression one can see that modes of the form
\begin{equation}\label{flat}
    ( \Xi^\mu,\, 0,\,  0,\,  0)  \qquad \text{such that} \qquad \Xi^\mu \partial_\mu K\rvert_{\rm vac} = 0\, ,
\end{equation}
are zero modes of the Hessian. Since in the branch {\bf S1} $\partial_\mu K\rvert_{\rm vac} \propto h_\mu$, such zero modes correspond to axionic modes of the the complex structure moduli that do not appear in the superpotential \eqref{WQ}. In fact, one can easily see that such directions do not appear in \eqref{VF}, and therefore are flat directions of the potential. 

In the following we will analyse further specific properties of {\bf H} for the branches \textbf{A1-S1} and \textbf{A2-S1}. For the former we will compute the mass spectrum for canonically normalised fields, finding that all tachyons satisfy the BF bound. For the latter we will directly show that {\bf H} is positive semidefinite, and therefore it contains no tachyons. Instead of tachyons, we will see that it contains additional zero modes compared to the other branches, in such a way that massless modes arrange into complex scalars. 

\subsection{Branch \textbf{A2-S1}}
\label{ap:ha2s1}

Let us first consider the Hessian in the branch \textbf{A2-S1} and, as stated above, show that it is positive semidefinite. By Sylverster's law of inertia, for showing that one may consider {\bf H} in any basis, without the need to express it in the basis of canonically normalised fields. Consider the expression \eqref{ap:Hfin}, which in the case at hand reads:
\begin{align}
\label{HA2S1}
{\bf H}\rvert_{\rm A2-S1}=e^K\tr^2\left(\begin{matrix}
 \frac{1}{18}\mk^2\p_\nu K\p_\mu K& -\frac{1}{6}\mk\partial_\nu K\mk_a& 0& 0\\
  \\
-\frac{1}{6}\mk\partial_\nu K\mk_a &-\frac{7}{3}\mk\mk_{ab}+\frac{7}{2}\mk_a\mk_b  &0 & \frac{4B}{3}\mk \mk_{ab}\\
  \\
0 & 0&\frac{1}{18}\mk^2\p_\alpha K	\p_\beta K & \frac{1}{6}\mk\partial_\alpha K\mk_b\\
  \\
 0& \frac{4B}{3}\mk \mk_{ab}&\frac{1}{6}\mk\partial_\alpha K\mk_b &-\frac{4}{3}\mk\mk_{ab}+\frac{7}{2}\mk_a\mk_b  \\
 \end{matrix}\right)\ , 
\end{align}
with $B=\pm1/2$.

Now, any (real) positive semidefinite matrix is a $n\times n$  symmetric matrix $M$ such that, for all non-zero  $x$ in $\mathds{R}^n$ satisfies $x^T Mx \geq0$. If one decomposes it as $M=\sum_i M_i$, and ech of the components satisfy
\begin{align}
x^T M_ix \geq0
\end{align} 
then it is straigtforard to see that
\begin{align}
x^T &M_ix \geq0 \quad \longrightarrow \quad x^TMx=x^T\sum_i M_ix\geq0 \, ,
\end{align}
which proves that $M$ is positive semidefinite. 

In the following we will use this property to show that \eqref{HA2S1} is positive semidefinite. We first decompose \eqref{HA2S1}  as
\begin{align}
    {\bf H}\rvert_{\rm A2-S1}= e^K\tr^2 \left({\bf X} + {\bf Y} + {\bf Z}\right)\,,
\end{align}
where
\begin{align}
{\bf X} & = 
\left(\begin{matrix}
 \frac{1}{18}\mk^2\p_\nu K\p_\mu K& -\frac{1}{6}\mk\partial_\nu K\mk_a& 0& 0\nonumber\\
  \\
-\frac{1}{6}\mk\partial_\nu K\mk_a &\frac{1}{2}\mk_a\mk_b  &0 & 0\\
  \\
0 & 0&0 & 0\\
  \\
0 & 0&0 & 0\\
 \end{matrix}\right)\,,
 {\bf Y}  = 
 \left(\begin{matrix}
0 & 0&0 & 0\\  \\
0 & 0&0 & 0\\
  \\
0 & 0&\frac{1}{18}\mk^2\p_\alpha K	\p_\beta K & \frac{1}{6}\mk\partial_\alpha K\mk_b\\
  \\
 0& 0&\frac{1}{6}\mk\partial_\alpha K\mk_b &\frac{1}{2}\mk_a\mk_b \\
 \end{matrix}\right)\, ,
\end{align}
and
\begin{align}
{\bf Z} & = 
\left(\begin{matrix}
0 & 0&0 & 0\\  \\
0 & -\frac{7}{3}\mk\mk_{ab}+3\mk_a\mk_b &0 & \frac{4B}{3}\mk \mk_{ab}\\
  \\
0 & 0&0 & 0\\
  \\
 0& \frac{4B}{3}\mk \mk_{ab}&0 &-\frac{4}{3}\mk\mk_{ab}+3\mk_a\mk_b  \\
 \end{matrix}\right)\, .\nonumber
\end{align}
We need to prove that each of these three matrices is positive semidefinite. Starting with {\bf X}, one can see that the non-trivial block can be decomposed as the following product 
\begin{align}
&\left(\begin{matrix}
 \frac{1}{18}\mk^2\p_\nu K\p_\mu K& -\frac{1}{6}\mk\partial_\mu K\mk_b\\
  \\
-\frac{1}{6}\mk\partial_\nu K\mk_a &\frac{1}{2}\mk_a\mk_b \\
 \end{matrix}\right)=\left(\begin{matrix}
 \frac{\sqrt{2}}{12}\mk\p_\mu K& 0\\
  \\
-\frac{\sqrt{2}}{4}\mk_a &0\\
 \end{matrix}\right)\left(\begin{matrix}
 4& 0\\
  \\
0 &K^{ab}\\
 \end{matrix}\right)\left(\begin{matrix}
 \frac{\sqrt{2}}{12}\mk\p_\nu K& -\frac{\sqrt{2}}{4}\mk_b\\
  \\
0 &0\\
 \end{matrix}\right)\, .
\end{align}
That is, it can be written as a Gramian matrix, which implies its positive-semidefiniteness. The same statement applies to the non-trivial block of the matrix {\bf Y}, which reads
\begin{align}
&\left(\begin{matrix}
 \frac{1}{18}\mk^2\p_\nu K	\p_\mu K & \frac{1}{6}\mk\partial_\mu K\mk_b\\
  \\
\frac{1}{6}\mk\partial_\nu K\mk_a &\frac{1}{2}\mk_a\mk_b  \\
 \end{matrix}\right)=\left(\begin{matrix}
 \frac{\sqrt{2}}{12}\mk\p_\mu K& 0\\
  \\
\frac{\sqrt{2}}{4}\mk_a &0\\
 \end{matrix}\right)\left(\begin{matrix}
 4& 0\\
  \\
0 &K^{ab}\\
 \end{matrix}\right)\left(\begin{matrix}
 \frac{\sqrt{2}}{12}\mk\p_\nu K& \frac{\sqrt{2}}{4}\mk_b\\
  \\
0 &0\\
 \end{matrix}\right)\, .
\end{align}

Things are slightly more involved for the non-trivial block of the matrix {\bf Z}. This reads
\begin{align}
\left(\begin{matrix}
-\frac{7}{3}\mk\mk_{ab}+3\mk_a\mk_b  & \pm \frac{2}{3}\mk \mk_{ab}\\
  \\
\pm \frac{2}{3}\mk \mk_{ab} &-\frac{4}{3}\mk\mk_{ab}+3\mk_a\mk_b  \\
 \end{matrix}\right)\, ,
\label{Zblock}
\end{align}
where we have considered for both choices of sign in $B=\pm 1/2$. In this case one can rewrite \eqref{Zblock} as:
\begin{align}
&\frac{2}{3} \left(\mk_a\mk_b-\mk \mk_{ab}\right)
\left(\begin{matrix}
1 & \mp 1 \\
  \\
\mp 1 & 1  \\
 \end{matrix}\right) +\left(\begin{matrix}
-\frac{5}{3}\mk\mk_{ab}+\frac{7}{3}\mk_a\mk_b  & \pm \frac{2}{3}\mk_{a}\mk_{b}\\
  \\
\pm \frac{2}{3}\mk_{a}\mk_{b} &-\frac{2}{3}\mk\mk_{ab}+\frac{7}{3}\mk_a\mk_b  \\
 \end{matrix}\right)\, .
\end{align}
The first matrix is a tensor product of two positive semidefinite matrices. The second one satisfies:
\begin{align}
&\left(\begin{matrix}
q^b	&	p^b
\end{matrix}\right)\left(\begin{matrix}
-\frac{5}{3}\mk\mk_{ab}+\frac{7}{3}\mk_a\mk_b  & \pm \frac{2}{3}\mk_{a}\mk_{b}\\
  \\
\pm \frac{2}{3}\mk_{a}\mk_{b} &-\frac{2}{3}\mk\mk_{ab}+\frac{7}{3}\mk_a\mk_b  \\
 \end{matrix}\right)\left(\begin{matrix}
q^a	\\
	p^a
\end{matrix}\right)\nonumber\\&=-\frac{5}{3}\mk\mk_{ab}q^aq^b+\frac{7}{3}\mk_a\mk_bq^aq^b \pm \frac{4}{3}\mk_{a}\mk_{b}q^ap^b-\frac{2}{3}\mk\mk_{ab}p^ap^b+\frac{7}{3}\mk_a\mk_bp^ap^b\nonumber\\ 
&=\frac{2}{3}\mk_a\mk_b \left(q^a \pm p^a\right)\left(q^b\pm p^b\right) +\frac{5}{3}\left(\mk_a\mk_b-\mk\mk_{ab}\right)q^aq^b +\frac{2}{3}\left(\mk_a\mk_b-\mk\mk_{ab}\right)p^ap^b+\mk_a\mk_b p^a p^b \geq 0
\end{align}
where we have used that all the metrics involved are positive semidefinite. Therefore {\bf Z} is also positive semidefinite.

Notice that the Hessian matrix \eqref{HA2S1} has further zero modes beyond the ones corresponding to the flat directions \eqref{flat}. These are of the form
\begin{equation}\label{zm}
    ( 0, \, 0,\,  \Xi^\mu,\,  0)  \qquad \text{such that} \qquad \Xi^\mu \partial_\mu K\rvert_{\rm vac} = 0\, ,
\end{equation}
and are nothing but the complex structure saxions that pair up with the axionic flat directions into complex scalar field. This time, as these fields appear in the potential via \eqref{ZAB}, they will not be flat directions of the potential. One can check that they develop a quartic potential, as discussed in section \ref{ap:complex} below.

\subsection{Branch \textbf{A1-S1}}\label{ap:ha1s1}

In this branch the Hessian \eqref{ap:Hfin} takes a block-diagonal form, namely
\begin{align}
    {\bf H}\rvert_{\rm A1-S1}= e^K\tr^2 
    \left(\begin{matrix}
    {\bf A} & 0 \\ 0 & {\bf S}\\
     \end{matrix}\right)\,,
\end{align}
where
\begin{align}\label{AxiM}
{\bf A} & = 
\left(\begin{matrix}
 \frac{8}{225}\mk^2\p_\nu K\p_\mu K& \frac{8C}{15}\mk\partial_\nu K\mk_a& \\
  \\
\frac{8C}{15}\mk\partial_\nu K\mk_a &\left(-\frac{4}{3}+\frac{8C}{3}\right)\mk\mk_{ab}+\frac{68}{25}\mk_a\mk_b \\
 \end{matrix}\right)\,,
 \end{align}
 \begin{align}\label{SaxiM}
 {\bf S}  = 
 \left(\begin{matrix}
\mk^2\left(\frac{4}{75}\p_\alpha K	\p_\beta K-\frac{16}{225} K_{\alpha\beta}\right) & \frac{8}{75}\mk\partial_\alpha K\mk_b\\
  \\
\frac{8}{75}\mk\partial_\alpha K\mk_b &-\frac{24}{25}\mk\mk_{ab}+\frac{68}{25}\mk_a\mk_b  \\
 \end{matrix}\right)\, ,
\end{align}
with $C=\pm 3/10$. Therefore, one can analyse the spectrum of axions or saxions separately. 


\subsubsection*{Axionic sector}

Let us first analyse the axionic sector. One can rewrite {\bf A} as:
\begin{align}
{\bf A}&=\left(\begin{matrix}
 \frac{\sqrt{2}}{15}\mk\p_\mu K& 0\\
  \\
C\sqrt{2}\mk_a &0\\
 \end{matrix}\right)\left(\begin{matrix}
 4& 0\\
  \\
0 &K^{ab}\\
 \end{matrix}\right)\left(\begin{matrix}
 \frac{\sqrt{2}}{15}\mk\p_\nu K& C\sqrt{2}\mk_b\\
  \\
0 &0\\
 \end{matrix}\right)+\nonumber
 \\&+\left(\begin{matrix}
 0&0\\
  \\
0 &\left(\frac{4}{3}-\frac{8C}{3}\right)\left(\mk_a\mk_b-\mk\mk_{ab}\right)+\left(\frac{2+8C}{3}\right)\mk_a\mk_b\\
 \end{matrix}\right)\, ,
 \label{tachC}
\end{align}
so for $C=\frac{3}{10}$ (i.e., the supersymmetric branch) the matrix {\bf A} \eqref{tachC} is a sum of positive semidefinite matrices, whereas for $C=-\frac{3}{10}$ the second one is not positive semidefinite. 

In order to compute the physical mass spectrum we need to express the Hessian in a basis of canonically normalised fields. For this, notice that the K\"ahler metrics for the K\"ahler and complex structure fields can be decomposed as:
\begin{align}
&K_{ab}=\frac{3}{2\mathcal{K}}\left(\frac{3\mathcal{K}_a\mathcal{K}_b}{2\mathcal{K}}-\mathcal{K}_{ab}\right)=\frac{3}{4}\frac{\mk_a\mk_b}{\mk^2} + \frac{3}{2\mk}\left(\frac{\mk_a\mk_b}{\mk}-\mk_{ab}\right)=K_{ab}^{\rm NP}+K_{ab}^{\rm P}\, ,\\
&K_{\mu\nu}=\frac{1}{16}\frac{\p_\mu G\p_\nu G}{G^2} + \frac{1}{4}\left(\frac{3}{4}\frac{\p_\mu G\p_\nu G}{G^2}-\frac{\p_\mu\p_\nu G}{G}\right)=K_{\mu\nu}^{\rm NP}+K_{\mu\nu}^{\rm P}\, ,
\end{align}
with $G = {\cal G}_T^2$, as defined below \eqref{KQ}. Here $K_{ab}^{\rm P}$ and $K_{ab}^{\rm NP}$ stand for the primitive and non-primitive factors of the K\"ahler moduli metric, which act on orthogonal subspaces of dimension $h^{1,1}_--1$ and 1. A similar decomposition holds for the metric of the dilaton-complex structure sector, now acting on spaces of dimension $N$ and 1, with $N$ the number of complex structure moduli. In terms of this decomposition, the matrix {\bf A} in the non-SUSY branch $C=-\frac{3}{10}$ reads
\begin{align}\label{AnonSUSY}
{\bf A} &= \left(\begin{matrix}
 \frac{128}{225}\mk^2K_{\mu\nu}^{\rm NP}& -\frac{4}{25}\mk\p_\nu K\mk_a\\
  \\
-\frac{4}{25}\mk\p_\nu K\mk_a& \frac{64}{45}\mk^2 K^{\rm P}_{ab}+\frac{176}{225}\mk^2 K_{ab}^{\rm NP}\\
 \end{matrix}\right)\, .
\end{align}

Now, the effective Lagrangian describing the axion spectrum will be of the form
\begin{equation}
L\supset\left(\p \xi^\mu \ \p b^a\right)\left( \begin{matrix}
K_{\mu\nu}\rvert_{\rm vac} &	0\\
0	&	K_{ab}\rvert_{\rm vac}
\end{matrix}\right) \left(\begin{matrix}\p \xi^\nu \\ \p b^b \end{matrix}\right)
+\frac{1}{2}\left( \xi^\nu \  b^a\right) \left[e^K \tr^2 {\bf A}\right]_{\rm vac} \left(\begin{matrix} \xi^\nu \\  b^b \end{matrix}\right)\, ,
\end{equation}
with {\bf A} given by \eqref{AnonSUSY} in the non-supersymmetric case. One can now define a basis of canonically normalised fields by performing the change of basis
\begin{equation}\label{canon}
    ( \xi^\mu \quad b^a) \quad \longrightarrow \quad  (\hat{\xi}\quad \hat{b} \quad \xi^{\hat{\mu}} \quad b^{\hat{a}} )\, ,
\end{equation}
where $\hat{b}$ is the vector along the subspace corresponding to $K_{ab}^{\rm NP}\rvert_{\rm vac}$, with unit norm, and similarly for $\hat{\xi}$ with $K_{\mu\nu}^{\rm NP}\rvert_{\rm vac}$. Finally, $\xi^{\hat{\mu}}$ with $\hat{\mu} = 1, \dots, N$ and $b^{\hat{a}}$ with $\hat{a} = 1, \dots h_-^{1,1} -1$ correspond to vectors of unit norm with respect to $K_{\mu\nu}^{\rm P}\rvert_{\rm vac}$ and $K_{ab}^{\rm P}\rvert_{\rm vac}$, respectively. One can see that in this new basis {\bf A} has the form
\begin{align}
    \hat{\bf A} & = \frac{16}{5}
    \left( \begin{matrix}
    \frac{8}{45} & \frac{2}{5\sqrt{3}} & & \\
    \frac{2}{5\sqrt{3}} & \frac{11}{45} & & \\
    & & 0 & \\
    & & & \frac{4}{9}
    \end{matrix}\right) \mk^2\, ,
\end{align}
and so the Hessian eigenvalues in the canonically normalised basis are
\begin{align}
   e^K \cK^2 \tr^2 \frac{8}{45} \left\{-\frac{1}{5}\, , \quad  4\, , \quad 0 \, ,  \quad 4\right\}\, .
\end{align}
Finally, one must compare such masses with the BF bound
\begin{equation}
    |m_{\rm BF}|^2 \, =\, - \frac{3}{4} V_{\rm vac} \, =\, e^K \frac{\CK^2\tr^2}{25}\, .
\end{equation}
In term of it one finds that the spectrum reads
\begin{align}
  m^2 =   \left\{-\frac{8}{9}\, ,  \quad  \frac{160}{9}\, , \quad 0  \, , \quad  \frac{160}{9}\right\}  |m_{\rm BF}|^2 \, ,
\end{align}
and so the tachyon in this sector does not induce an instability. 

For completeness, let us finish this section by computing also the spectrum for the SUSY case.  Proceeding exactly as before but taking $C=\frac{3}{10}$ it is straightforward to obtain the following eigenvalues for the canonically normalised Hessian:
\begin{align}
   e^K \cK^2 \tr^2 \frac{8}{45} \left\{\frac{44}{5}\, , \quad   1\, , \quad 0 \, ,  \quad 1\right\}\, ,
\end{align}
or in terms of the BF bound:
\begin{align}
  m^2 =   \left\{\frac{352}{9}\, ,  \quad  \frac{40}{9}\, , \quad 0  \, , \quad  \frac{40}{9}\right\}  |m_{\rm BF}|^2 \, .
\end{align}
\subsubsection*{Saxionic sector}

Let us now analyse the spectrum in the saxionic sector. Notice that this time the matrix \eqref{SaxiM} is independent of the sign of $C$, and so the tachyonic directions that one may find will be common to the supersymmetric and non-supersymmetric branches of the kind {\bf A1-S1}. Since the supersymmetric branch should not contain any classical instability, neither should there be one for its non-supersymmetric counterpart. Let us nevertheless confirm this expectation explicitly. 

As before we first rewrite \eqref{SaxiM} as
\begin{align}
\label{Snew}
{\bf S} &=\left(\begin{matrix}
\frac{176}{225}\mk^2K_{\mu\nu}^{\rm NP}-\frac{16}{225}\mk^2K^{\rm P}_{\mu\nu}	&	\frac{8}{75}\mk\partial_\alpha K\mk_b\\
  \\
\frac{8}{75}\mk\partial_\alpha K\mk_b & \frac{48}{75} \cK^2 K^{\rm P}_{ab} + \frac{176}{75} \cK^2 K^{\rm NP}_{ab}  \\
 \end{matrix}\right)\, .
\end{align}
Then we perform a change of basis for the saxions
\begin{equation}
    ( u^\mu \quad t^a) \quad \longrightarrow \quad  (\hat{u}\quad \hat{t} \quad u^{\hat{\mu}} \quad t^{\hat{a}} )\, ,
\end{equation}
with analogous definitions as in \eqref{canon}. In this basis the matrix {\bf S} reads
\begin{align}
    \hat{\bf S} & = \frac{16}{75}
    \left( \begin{matrix}
    \frac{11}{3} & -\frac{4}{\sqrt{3}} & & \\
    -\frac{4}{\sqrt{3}} & 11 & & \\
    & & - \frac{1}{3} & \\
    & & & 3
    \end{matrix}\right) \mk^2\, ,
\end{align}
and so the Hessian eigenvalues in the canonically normalised basis are
\begin{align}
   e^K \cK^2 \tr^2 \frac{8}{75} \left\{3 \, ,  \quad \frac{35}{3}\, ,\quad  -\frac{1}{3}\, , \quad  3 \right\}\, ,
\end{align}
where now the tachyonic eigenvalue has a degeneracy of $N$, as it corresponds to the `primitive' complex structure saxions $u^{\hat{\mu}}$. Comparing with the BF bound one finds 
\begin{align}
  m^2 =   \left\{8 \, , \quad  \frac{280}{9}\, , \quad  -\frac{8}{9}\, ,  \quad  8 \right\}  |m_{\rm BF}|^2 \, .
\end{align}
As expected, the tachyonic directions in this sector do not induce a classical instability. 
\subsection{Complex structure saxions}
\label{ap:complex}

In the superpotential \eqref{WQ} only one linear combination of dilaton and complex structure moduli appear. As a direct consequence we have $N$ axionic flat directions of the potential, where $N$ is the number of complex structure moduli. In the following we would like to analyse the potential that it is induced for their saxionic partners. This question is particularly relevant for the branch {\bf A2-S1} of vacua, where such saxionic modes are found to be massless.

Let us consider the linear combinations of complex structure and dilaton moduli $U^i = \xi^i + i u^i$ not appearing in the superpotential \eqref{WQ}. Then, one can check that they satisfy the property
\begin{equation}
\left[\p_{u^i} K\right]_{\rm vac}=0\, .
\end{equation}
Using this it is straightforward to see that at the vacuum
\begin{align}
\p_{u^i} V\rvert_{\rm vac} & = \left[e^K \left(\p_{u^i} K \tilde{V}+ e^K\p_{u^i}\tilde{V}\right)\right]_{\rm vac}=0\, , \\
\p_{u^i}\p_{u^j} V\rvert_{\rm vac} & = \left[e^K\left(\p_{u^i}\p_{u^j} K \tilde{V}+\p_{u^i}\p_{u^j} \tilde{V}\right)\right]_{\rm vac} = {\cal A} \left[e^K\p_{u^i}\p_{u^j} K\right]_{\rm vac}\, ,
\end{align}
where we have defined  $\tilde{V}=e^{-K}V$ and 
\begin{align}
{\cal A} = \left(8A^2-\frac{2B^2}{27}-\frac{16C^2}{27}\right)\, .
\label{combi}
\end{align}
Replacing the values for the constants $A, B, C$ for the different branches in the second equation, one recovers the corresponding sector of the Hessian. In particular, one can check that \eqref{combi} vanishes for the branch {\bf A2-S1}, as expected. 

One may then proceed and compute further derivatives of the potential at the vacuum:
\begin{align}
\p_{u^i}\p_{u^j}\p_{u^l} V\rvert_{\rm vac}&=\left[e^K\left(\p_{u^i}\p_{u^j}\p_{u^l} K \tilde{V}+\p_{u^l}\p_{u^i}\p_{u^j} \tilde{V}\right) \right]_{\rm vac} = {\cal A} \left[e^K\p_{u^i}\p_{u^l}\p_{u^j} K\right]_{\rm vac}\, ,\\
\p_{u^i}\p_{u^j}\p_{u^l}\p_{u^m} V\rvert_{\rm vac}&=128A^2 \left[e^K \left( K_{u^ju^l}K_{u^iu^m}+K_{u^ju^m}K_{u^iu^l}+K_{u^lu^m}K_{u^iu^j} \right)\right]_{\rm vac}+\dots
\end{align}
where the dots stand for terms proportional to ${\cal A}$. As the term in brackets is a product of kinetic terms, in the case ${\cal A} = 0$ we obtain a non-vanishing, positive quartic coupling. This completes the proof that the branch {\bf A2-S1} features a positive semidefinite potential in the vicinity of the vacuum.

\subsection{Adding mobile D6-branes}
\label{ap:openH}

In the presence of mobile D6-branes and for each extremum found in section \ref{s:D6branes}, one can show that the formalism developed in section \ref{ss:hessian} is still valid. The matrix of second derivatives takes the form: $$\p_\alpha\p_\beta V'\rvert_{\rm vac}=\p_\alpha\p_\beta \left(V_1'+V_2'\right)\rvert_{\rm vac}=2\left(\p_\alpha\vec{\g'}^t\right) \hat{\bf Z'}_{1} \left(\p_\beta\vec{\g'}+\vec{\eta'_\beta}\right)\, ,$$ with the correspondent redefinition of $\{\hat{\bf Z}'_1, \p_\alpha \vec{\gamma}', \eta_{\beta}'\}$ incorporating the open string moduli and $\{V_1'$, $V_2'\}$ introduced in \eqref{openpotential}. The matrix $\hat{\bf Z}_1'$ is defined, analogously to \eqref{Vsplit}, such that $V_1'=\vec{\gamma}^{t'}\hat{\bf Z}'_1\vec{\gamma}'$ is quadratic on quantities that vanish in the vacuum. Looking at \eqref{openpotential} it is straightforward to see that:
\begin{equation}
\hat{\bf Z}_1' = 
\left(
\begin{matrix}
 4 & 0 & 0 & 0 & 0 & 0\\
 \\
 0 & K^{ab} & 0 & 0 & 0 & 0 \\
 \\
 0 & 0 & \frac{4}{9} \mk^2 K_{ab} & 0 & 0 & 0 \\
 \\
  0 & 0 & 0 & K^{\mu \nu } & 0 & 0\\
 \\
  0 & 0 & 0 & 0 &  G^{ij} & 0 \\
  \\
0 & 0 & 0 & 0 & 0 & G^{ij}    \\
\end{matrix}
\right)\, .
\end{equation}
Regarding the new $\p_\alpha \vec{\gamma}'$'s its explicit expression can be computed directly from \eqref{gammaopen}:
\begin{align}
\partial_{\xi^\mu}\vec{\g}^t=&\left(\begin{matrix}
h_\mu ,&0 ,&0, &0,&0 ,&0 
\end{matrix}\right)\, ,\nonumber \\
\partial_{b^c}\vec{\g}^t=&\left(\begin{matrix}
\rho_c-\hth^i\r_{ci}, &\mk_{acd}\tr^d -f^i_a \rho_{c\, i},&\delta^a_c\tr ,&0, & -\rho_{c\, i},&0
\end{matrix}\right)\, ,\nonumber \\
\partial_{\hth^i}\vec{\g}^t=&\left(\begin{matrix}
\r_i, &-\r_{ai},&0 ,&0, & 0,&0
\end{matrix}\right)\, ,\nonumber \\
\partial_{u^\alpha}\vec{\g}^t=&\left(\begin{matrix}
0, &0, &0, &-\tr A\mk\p_\alpha\partial_\nu K-\tr\mk\p_\alpha\tilde{\epsilon}^p_\nu,&0 ,&0 
\end{matrix}\right)\, ,\nonumber \\
\partial_{t^c}\vec{\g}^t=&\left(\begin{matrix}
0, & -2\tr C\mk_{ac}, &-\tr B\delta^a_c,&-3\tr A\mk_c\partial_\nu K -3\tr\mk_c\tilde{\epsilon}^p_\mu,0,0
\end{matrix}\right)+\nonumber \\+&\left(\begin{matrix}
0, & \p_{t^c}\left(f^i_a\r_i-\frac{1}{2}H^\mu_a\hr_\mu\right), &-\p_{t^c}\left(\mk^{ab}\phi^i+\mk^{ad}t^bf^i_d\right)\r_{bi},&-\frac{1}{2}\p_{t^c}g^\mu_i\hr_\mu,&\r_{ci}
\end{matrix}\right)\, ,\nonumber\\
\partial_{\phi^i}\vec{\g}^t=&\left(\begin{matrix}
0,&\p_{\phi^i}\left(f^j_a\r_j-\frac{1}{2}H^\mu_a\hr_\mu\right),& -\mk^{ab}\r_{bi}-\mk^{ad}t^b\p_{\phi^i}f^j_d\r_{bj},&0
\end{matrix}\right)\, .
\end{align}
Finally, the  $\eta_{\alpha}'$'s are obatined by direct computation rewriting the second derivatives of $V'_2$  as:
\begin{equation}
\partial_\alpha \partial_\beta V'_2\rvert_{\rm vac}=2\vec{\eta'_\alpha}^t \hat{\bf Z'}_{1} \p_\beta\vec{\gamma'}= 2\p_\alpha\vec{\g'}^{\, t} \hat{\bf Z'}_{1} \vec{\eta'_\beta}\, .
\end{equation}
The  result is:
\begin{align}\nonumber
\vec{\eta}_{\xi^\mu}{}^t=& \left(\begin{matrix}
0,&0, &0, &0,	&	0,	&	0
\end{matrix}   \right)\, ,
\\
\vec{\eta}_{b^d}{}^t=&\tr\left(\begin{matrix}
0, &0,	&	\frac{3C}{\mk} K^{bc}\mk_{cd},	&	0,	&	0,	&	-\frac{4C}{3}\mk f^k_cG_{kj}
\end{matrix}\right)\, ,\nonumber \\ \nonumber
\vec{\eta}_{\hth^i}{}^t=& \tr\left(\begin{matrix}
0,&0, &0, &0,	&	0,	&	-\frac{4C}{3}\mk\frac{G_{ij}}{\delta^i_i}
\end{matrix}   \right)\, ,
\\
\vec{\eta}_{u^\alpha}{}^t=&\tr\left(\begin{matrix}
0, &C\p_\alpha K\mk_a,	&B\p_\alpha K t^a,	  	&	\left(\frac{2}{3}-4A\right)\mk\left(K_{\alpha\mu}-\frac{1}{4}\p_\mu K \p_\alpha K\right), &0, &0\end{matrix}\right)+ \nonumber
\\&+\tr\left(\begin{matrix}
0, &0,	&0,	  	&	e^{-K}\mk\tr K_{\beta\mu}\p_\alpha \left(e^KK^{\gamma\beta}\te^p_\gamma\right), &0, &0\end{matrix}\right)\nonumber \, ,
 \\
\vec{\eta}_{t^d}{}^t=&\tr\left(\begin{matrix}
0, & \frac{4C\mk}{3} K_{bd},	&	\frac{3B}{2\mk} K^{bc}\mk_{cd}, 	&	\mk\p_{t^d}\te^p_\mu, &-\frac{4C}{3}\mk^2 f^k_c G_{kj}, &\frac{2B}{3}\mk^2 f^k_c G_{kj}
\end{matrix}\right)+\frac{\left(H^\alpha_d-f^i_dg^\alpha_i\right)}{4} \vec{\eta_{u^\alpha}}^t\, ,\nonumber \\
\vec{\eta}_{\phi^i}{}^t=& \tr\left(\begin{matrix}
0,&0, &0, &0,	&\frac{4C}{3}\mk^2 \frac{G_{ij}}{\delta^i_i}, &-\frac{2B}{3}\mk^2 \frac{G_{ij}}{\delta^i_i}
\end{matrix}   \right)+\frac{1}{4}g^\alpha_i{\eta_{u^\alpha}}^t\, .
\end{align}


\bibliographystyle{JHEP2015}
\bibliography{refs_typeIIA}

\end{document}